\documentclass{article}
\usepackage{arxiv}
\usepackage{amsmath,amsfonts}
\usepackage{algorithmic}
\usepackage{algorithm}
\usepackage{array}
\usepackage[caption=false,font=normalsize,labelfont=sf,textfont=sf]{subfig}
\usepackage{textcomp}
\usepackage{hyperref}
\usepackage{dblfloatfix}
\usepackage{url}
\usepackage{verbatim}
\usepackage{graphicx}
\usepackage{cite}
\usepackage{breqn}
\usepackage{setspace}
\usepackage{geometry}
\usepackage[flushleft]{threeparttable}
\usepackage{graphicx}
\usepackage{tikz}
\usepackage{tabularx}

\newcommand{\subfigimg}[3][,]{%
	\setbox1=\hbox{\includegraphics[#1]{#3}}% Store image in box
	\leavevmode\rlap{\usebox1}% Print image
	\rlap{\hspace*{0pt}\raisebox{\dimexpr\ht1-1\baselineskip}{#2}}% Print label
	\phantom{\usebox1}% Insert appropriate spcing
}

\title{Ensemble Deep Learning for enhanced seismic data reconstruction}

\author{
	Mohammad Mahdi Abedi\textsuperscript{1},
	David Pardo\textsuperscript{2,1,3},
	Tariq Alkhalifah\textsuperscript{4}\\
	\textsuperscript{1}Basque Center for Applied Mathematics, Bilbao, Spain\\
	\textsuperscript{2}University of the Basque Country, Department of Mathematics, Spain\\
	\textsuperscript{3}Ikerbasque, Basque Foundation for Science, Bilbao, Spain\\
	\textsuperscript{4}King Abdullah University of Science and Technology, Thuwal 23955-6900, Saudi Arabia\\
	\textit{Emails:} mabedi@bcamath.org, david.pardo@ehu.es, tariq.alkhalifah@kaust.edu.sa
}

\begin{document}
	\bibliographystyle{unsrt}
	\maketitle
   \footnotetext{Under revision}
	
	\begin{abstract}
Seismic data often contain gaps due to various obstacles in the investigated area and recording instrument failures. Deep learning techniques offer promising solutions for reconstructing missing data parts by leveraging existing information. However, self-supervised methods frequently struggle with capturing under-represented features such as weaker events, crossing dips, and higher frequencies. To address these challenges, we propose a novel ensemble deep model along with a tailored self-supervised training approach for reconstructing seismic data with consecutive missing traces. Our model comprises two branches of U-nets, each fed from distinct data transformation modules aimed at amplifying under-represented features and promoting diversity among learners. Our loss function minimizes relative errors at the outputs of individual branches and the entire model, ensuring accurate reconstruction of various features while maintaining overall data integrity. Additionally, we employ masking while training to enhance sample diversity and memory efficiency. Application on two benchmark synthetic datasets and two real datasets demonstrates improved accuracy compared to a conventional U-net, successfully reconstructing weak events, diffractions, higher frequencies, and reflections obscured by groundroll. However, our method requires a threefold of training time compared to a simple U-net.
		
		
	\end{abstract}
	
	\section{Introduction}
	Seismic data often include recording gaps due to factors such as equipment limitations, irregular acquisition geometries, and strong noise. Deep learning has been applied to the reconstruction of these data gaps, leveraging neural networks' ability to learn complex patterns. Various studies have employed diverse deep-learning architectures, including autoencoders (\cite{mandelli2018seismic}), U-nets (\cite{park2019reconstruction, tang2020reconstruction}), long short-term memory networks (\cite{Yoon2021}), attention convolutional neural networks (CNNs) \cite{Li2022, Yu2022}, multistage CNN (\cite{he2022}), multiscale CNN (\cite{Cheng2023, Zhong2023}), multidirectional CNN (\cite{abedi2022multidirectional}), multi-cascade CNN (\cite{dong2024seismic}), diffusion probabilistic \cite{liu2024generative}, and other state-of-art models (e.g., \cite{Liu2023Consecutively,chen2023projection}).
	
	Traditional supervised methods for seismic data reconstruction require large and comprehensive datasets for training. Otherwise, they face considerable generalization issues when applied to realistic or target data (\cite{yu2021deep, safonova2023ten, zhao2023comparison}). On the other hand, state-of-the-art self-supervised methods learn inherent structures or relationships within a specific dataset to predict its gaps. Given the challenges in accessing globally representative seismic datasets due to significant variations in their acquisition parameters, processing techniques, and field parameters, self-supervised methods have recently gained attention in data reconstruction tasks (\cite{Wang2022, Meng2022, abedi2022multidirectional, liu2023trace, liu2024gabor, abedi2024semi}). In self-supervised seismic data reconstruction, the (often small) target dataset is utilized in both training and inference. However, due to the limited size of the available target data and the complexity of seismic structures, self-supervised deep-learning methods may struggle to learn some important, but poorly represented, features in the data. Reported instances of fine details challenging conventional methods include weaker events (e.g., \cite{he2022, Liu2023Consecutively}), crossing dips (e.g., \cite{brandolin2023pinnslope}), and high frequencies (e.g., \cite{huff2024near}). Improving the learnability of these fine details is the objective of this work, which introduces an ensemble framework to address this challenge.
	
	Ensemble deep learning involves training multiple models and combining their predictions to enhance the overall performance. It helps in improving the generalization and robustness, especially when a single model struggles with complex patterns or insufficient data samples (\cite{ganaie2022ensemble, mohammed2023comprehensive}). Combining predictions from multiple models also helps mitigate overfitting that is associated with small training datasets. The success of ensembles can be attributed to factors such as statistical, optimization, and representation learning (\cite{dietterich2000ensemble, ganaie2022ensemble}). Ensembles reduce statistical challenges arising from limited data, reduce the risk of being stuck into undesired local minima during optimization (e.g., by utilizing different starting weights in each model), and expand the representation space when individual models fall short (\cite{ganaie2022ensemble}).
	
	The effectiveness of ensemble models significantly relies on the diversity exhibited by the baseline learners within the ensemble (\cite{bian2007diversity}). Diverse perspectives and approaches among baseline models within an ensemble enhance their ability to generalize and make accurate predictions. Traditional methods for achieving diversity employ different training methods, data subsets, architectures, or initializations for each baseline (\cite{ganaie2022ensemble, mohammed2023comprehensive}). The multidirectional CNN (\cite{abedi2022multidirectional}) is an example of using ensemble deep learning for 3-D data reconstruction, using vertical and horizontal kernels to induce diversity in baseline models. 
	
To address the challenges posed by limited dataset size and enhance the model's ability to learn various features and generalize effectively within a given dataset, we propose an ensemble deep learning approach featuring specialized branches designed to learn specific aspects of the data. By incorporating specific data transformations within both the loss function and architecture, we train each baseline in the ensemble model on a subset or representation of the feature space, promoting diversity among baselines. We propose three types of transformations designed to amplify or attenuate events characterized by specific ranges of amplitudes, dips, and frequencies. In the implementation of our method in each example, we use  one type of the proposed transformations with two distinct hyperparameters to enable the baselines to complement or refine each other's predictions. Notably, our approach also offers a degree of explainability, as the influence of each baseline is determined by the design of the data transformations.
	
In the subsequent sections, we introduce our ensemble deep model (EDM), detailing both its loss function and architecture. Following this, we introduce three data transformations -- gamma correction, peak-frequency shifting, and dip filtering -- tailored for integration into our EDM framework. We explain our self-supervised training algorithm and showcase the application of our method using two benchmark synthetic datasets and two real seismic datasets. Through these demonstrations, we highlight the enhanced results of our approach compared to conventional models. Finally, we discuss various considerations regarding our proposed approach, including the additional hyperparameters within the loss function and transformations, the associated extra cost of our method, and potential paths for its extension.
	
\graphicspath{{./Figures/}}

	\section{Ensemble Deep model}
An Ensemble Deep Model (EDM) consists of multiple baseline models and a fusion component that combines the outcomes of the baseline models. Using a set of baseline deep models $\left\{h_1,\ldots,h_n\right\}$ and a fusion function $S$, we define an EDM as,
\begin{dmath}
\label{eq1}
H\left(d\right)=S\left(d,h_1\left(d\right),\ldots,h_n\left(d\right)\right),	
\end{dmath}
\noindent
where $H\left(d\right)$ is the ensemble and $h_j\left(d\right)$ is a baseline's reconstruction output for the input data sample $d$. 

To induce diversity among the baseline learners and enable the network to learn specific features, we use data transformations that intensify those features and employ them in both the loss function and architecture. We define each baseline $h_j\left(d\right)$ in the architecture as a composition of a data transformation $T_j\left(.\right)$, a U-Net model $U_{j}(.)$ \cite{ronneberger2015u}  , and the corresponding inverse transformation $T_j^{-1}\left(.\right)$:
\begin{dmath}
	\label{eq2}
	\begin{aligned}
		h_j(d) &= T_j^{-1}(U_j(T_j(d))), &\quad j &= 1,\ldots,n.
	\end{aligned}	
\end{dmath}
\noindent

We define the loss function as a weighted sum of the reconstruction errors of the ensemble and the output of each branch. For the case of two branches ($n=2$), we have:
\begin{dmath}
\label{eq3}
\mathcal{L}=E\left(H\left(d\right),d_{gt}\right)+\lambda_1E\left(U_1\left(T_1\left(d\right)\right),T_1\left(d_{gt}\right)\right)+\ \lambda_2E\left(U_2\left(T_2\left(d\right)\right),T_2\left(d_{gt}\right)\right), 
\end{dmath}

where, $d_{gt}$ is the ground truth, $\lambda_j$ are weight numbers, and $E$ represents the relative mean absolute error, defined as:

\begin{dmath}
\label{eq4}
E\left(\widetilde{y},y\right)=\frac{\sum_{i}\left|y_i-\widetilde{y_i}\right|}{\sum_{i}\left|y_i\right|} .
\end{dmath}

The summations in equation \ref{eq4} are over $i$, the index of elements in data arrays $\widetilde{y}$ and $y$ (represent the first and second terms in parenthesis in front of $E$ in equation \ref{eq3}). Utilizing the relative errors rather than absolute ones serves to compensate for amplitude variations of the various transformations used for each baseline, ensuring comparability among the contributions of various terms in the loss function.

\begin{figure*}[]
		\centering
		\includegraphics[width=.3\linewidth]{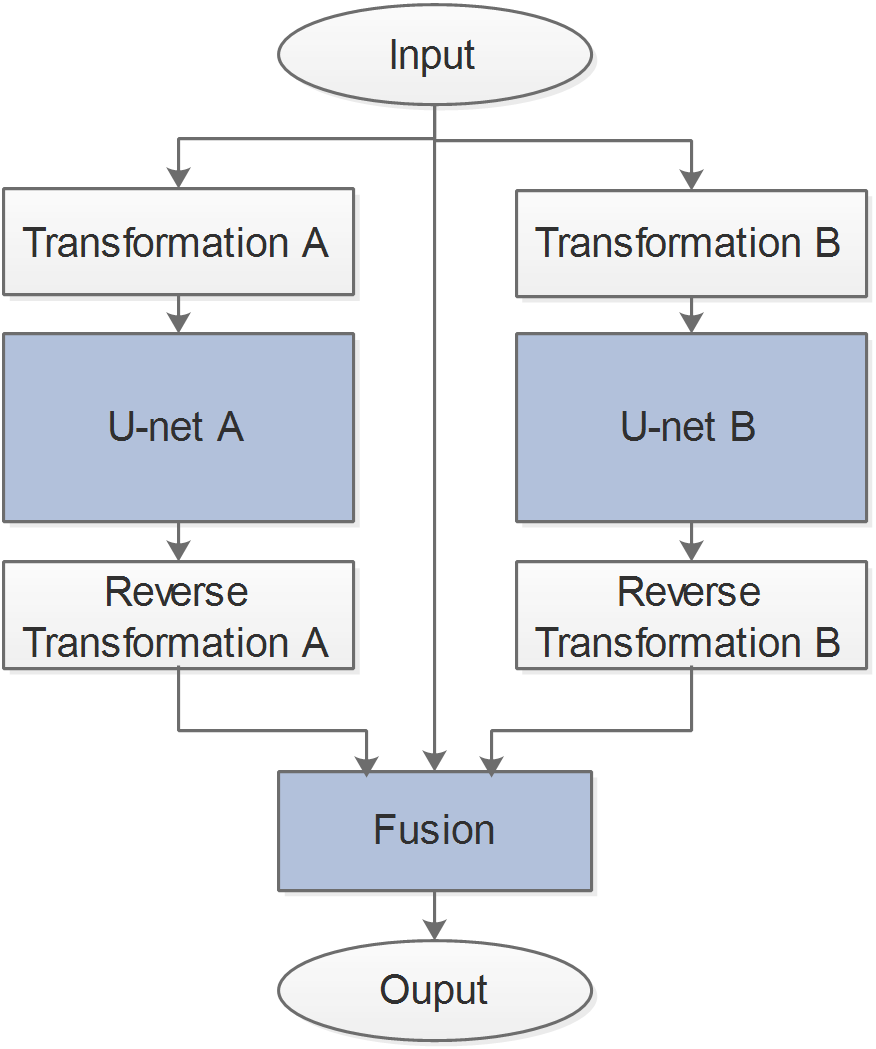}
		\caption{A diagram of the proposed ensemble deep model. The darker boxes include the trainable layers. The entire model is trained using as the loss function the weighted sum of errors at the outputs of U-net A, U-net B, as well as the Fusion block. Details of the networks are presented in Table 1.}
		\label{fig1}
\end{figure*}

Figure \ref{fig1} shows a sketch of the proposed network. During training and inference, each baseline model is fed with a differently transformed representation of the dataset, while the fusion part combines the best predictions in the target data domain. Employing the transformation modules, the domain changes are performed explicitly, thereby constraining the network's learning task solely to data reconstruction. 

Our method employs a parallel stacking ensemble structure, consisting of two branches of baseline models and a small convolutional network for fusion. This configuration is recognized as a bias-reducing technique (\cite{ganaie2022ensemble}) and is also used in (\cite{abedi2022multidirectional}). Each baseline model's trainable parameters are part of a U-net architecture (\cite{ronneberger2015u}), commonly applied in data reconstruction tasks. Table 1 shows the architectural details. Four strided convolutional layers are used for downsampling in the encoder, followed by the corresponding upsampling with four strided transpose convolutional layers in the decoder. Skip connections facilitate the transfer of low-level information to later layers, enhancing the network's performance. The employed down/upsampling strategy increases the network's receptive field. To accommodate the higher temporal resolution of seismic data compared to the horizontal resolution, we implement a larger vertical stride in the first and last layers, aiding the network in effectively capturing steeper seismic events. The selection of a CNN for the fusion process is more effective than employing a simplistic averaging of model outputs as averaging lacks data adaptability and is susceptible to biased learners (\cite{ju2018relative}).

	\begin{table}
		\centering
		\begin{threeparttable}
			
			\caption{Detailed architecture of the trainable blocks of the proposed EDM (see Figure \ref{fig1}).  \label{tab:table1}}
\begin{tabularx}{.7\textwidth}{lll}
	\hline
   & \textbf{Layers} & \textbf{Data dimensions} \\
	\hline
	\rotatebox[origin=c]{90}{\textbf{U-net }}&1. Input & $(H, W, 1)$ \\
	&2. Conv2D (5x5,4x2), ReLU & $(H/4, W/2, 32)$ \\
	&3. Conv2D (3x3,2x2), BN, ReLU & $(H/8, W/4, 64)$ \\
	&4. Conv2D (3x3,2x2), BN, ReLU & $(H/16, W/8, 128)$ \\
	&5. Conv2D (3x3,2x2), BN, ReLU & $(H/32, W/16, 128)$ \\
	&6. TrConv2D (3x3,2x2), BN, ReLU & $(H/16, W/8, 128)$ \\
	&7. Concat (4,6), TrConv2D (3x3,2x2), BN, ReLU & $(H/8, W/4, 64)$ \\
	&8. Concat (3,7), TrConv2D (3x3,2x2), BN, ReLU & $(H/4, W/2, 32)$ \\
	&9. Concat (2,8), TrConv2D (5x5,4x2) & $(H, W, 1)$ \\
	\hline
	\hline
	\rotatebox[origin=c]{90}{\textbf{Fusion }}&1. Input & $(H, W, 3)$ \\
	&2. Conv2D (3x3,2x2), BN, ReLU & $(H/2, W/2, 64)$ \\
	&3. Conv2D (3x3,1x1), BN, ReLU & $(H/2, W/2, 32)$ \\
	&4. TrConv2D (3x3,2x2) & $(H, W, 1)$ \\
	\hline
\end{tabularx}
			\begin{tablenotes}
				\item We use 2D convolution (Conv2D), Transpose convolution (TrConv2D), Batch normalization (BN), and Concatenation (Concat) layers, besides the ReLU activation function. In parentheses besides the Conv2D and TrConv2D, we show the kernel and strides size. H and W are the height and width of the input image.
			\end{tablenotes}
		\end{threeparttable}
	\end{table}

\begin{figure}[]
	\centering
	\begin{tabular}{@{}p{\linewidth}@{\quad}p{\linewidth}@{}}
		\centering
		\subfigimg[height=6cm]{(a)}{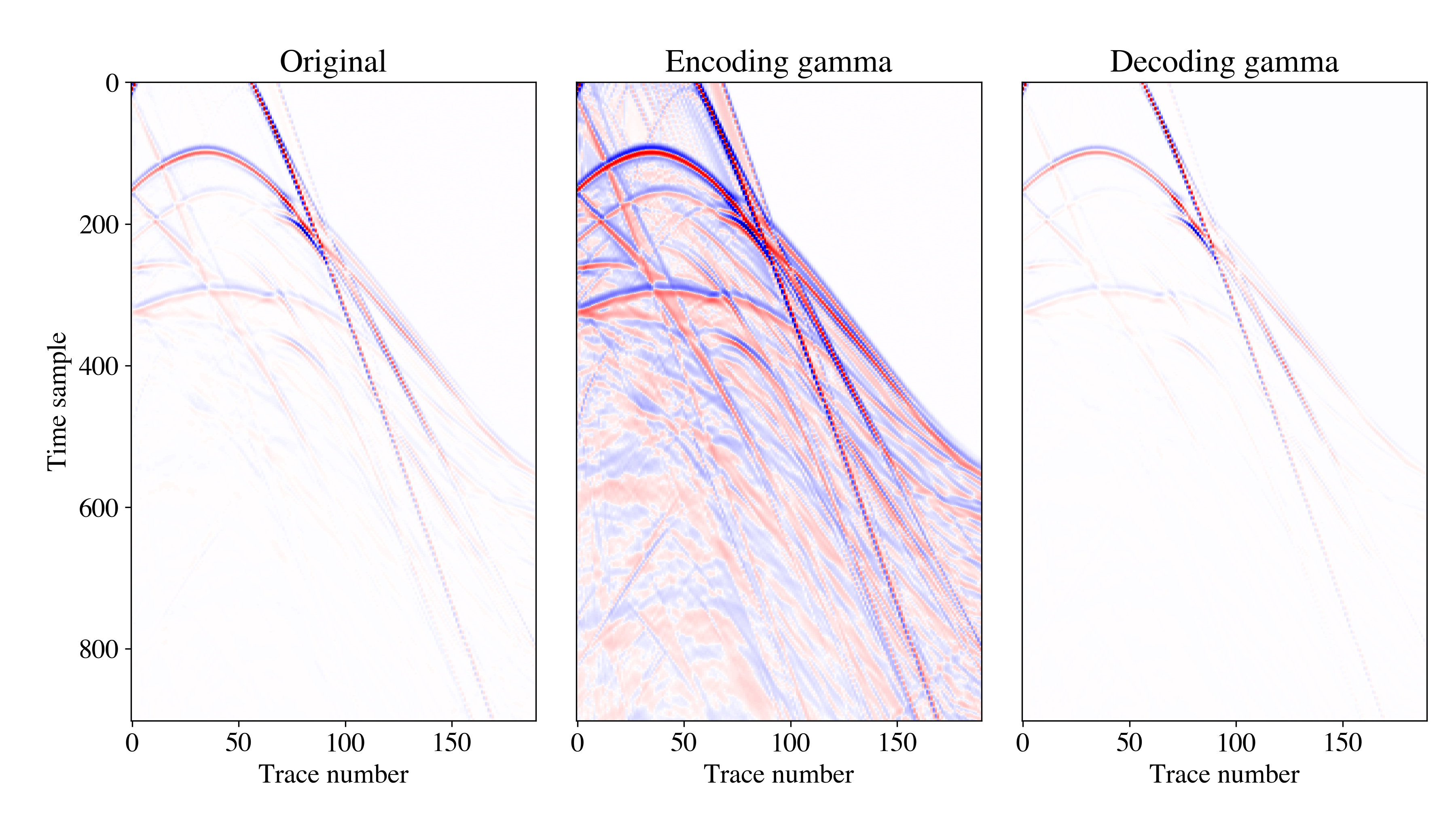} 
		\subfigimg[height=6cm]{(b)}{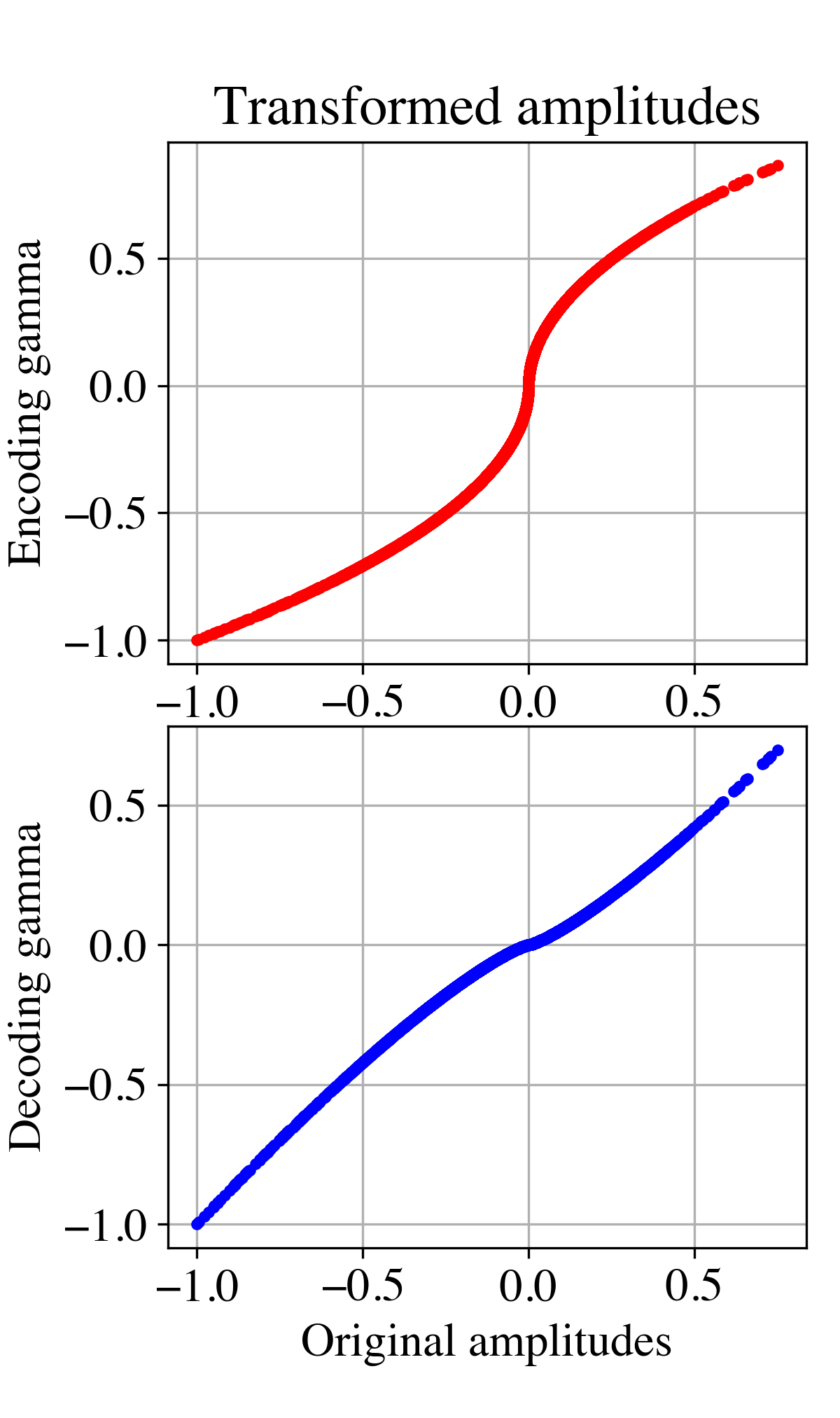} 
		\\
		\subfigimg[height=6cm]{(c)}{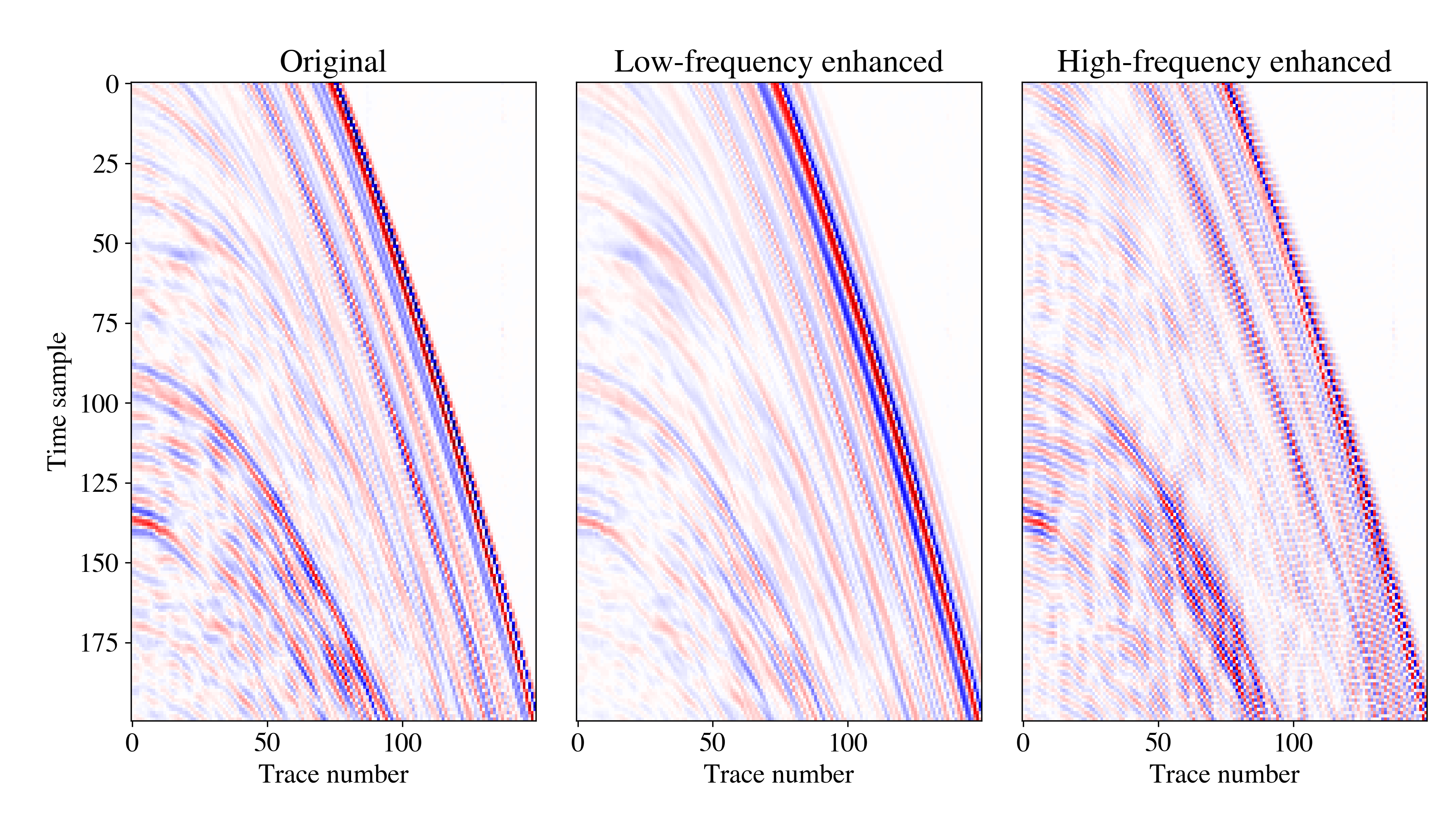} 
		\subfigimg[height=6cm]{(d)}{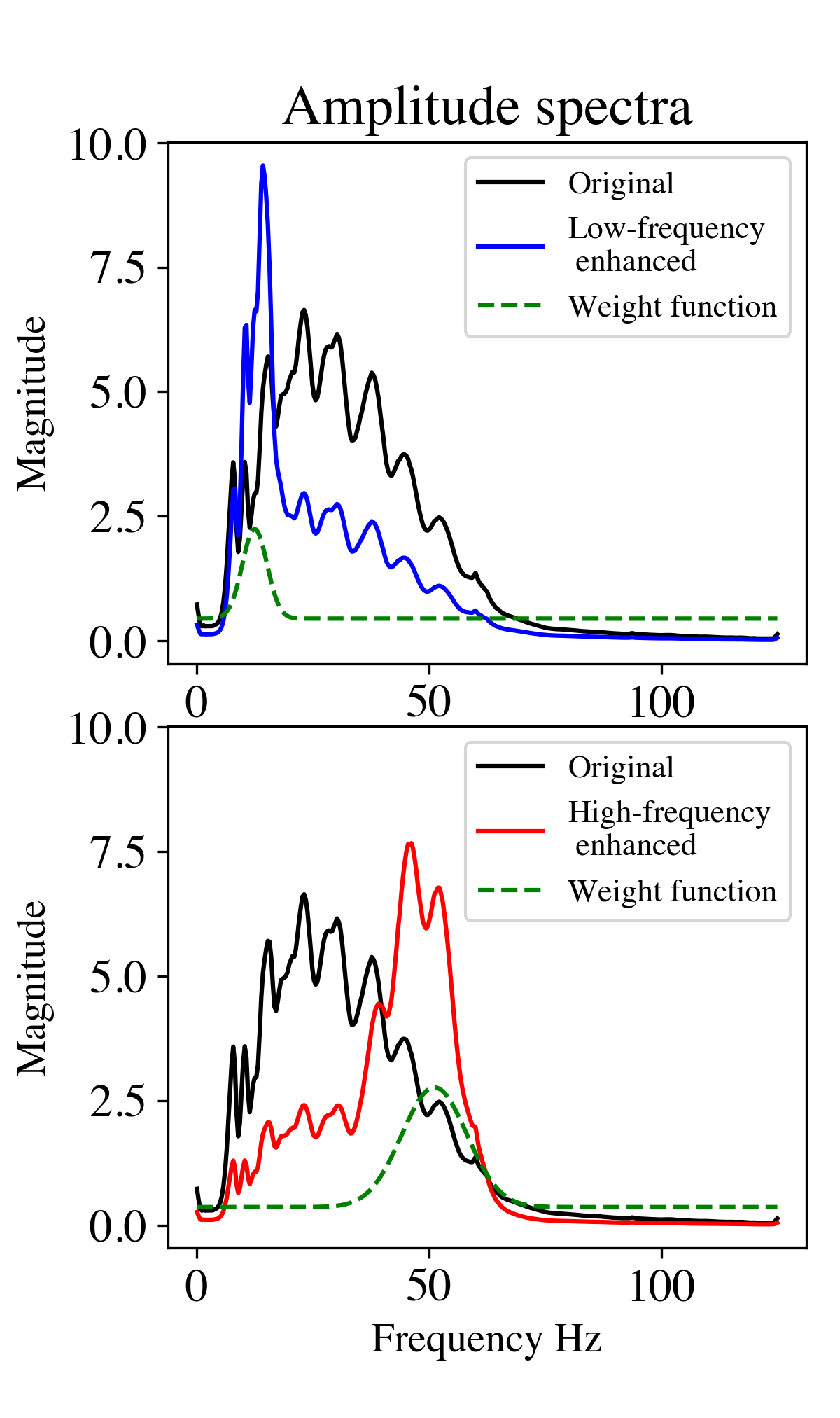} 
		\\
		\subfigimg[height=6cm]{(e)}{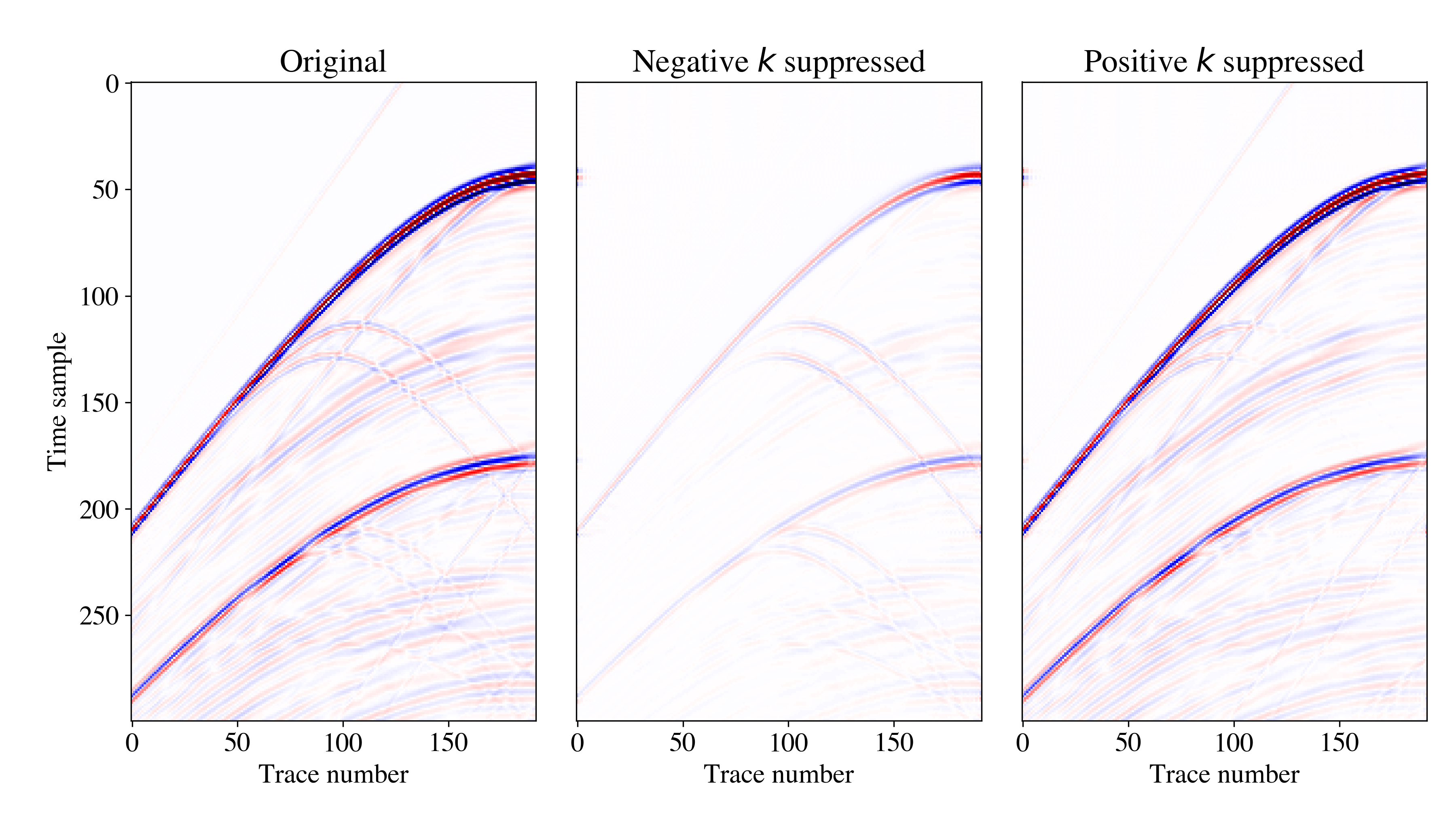} 
		\subfigimg[height=6cm]{(f)}{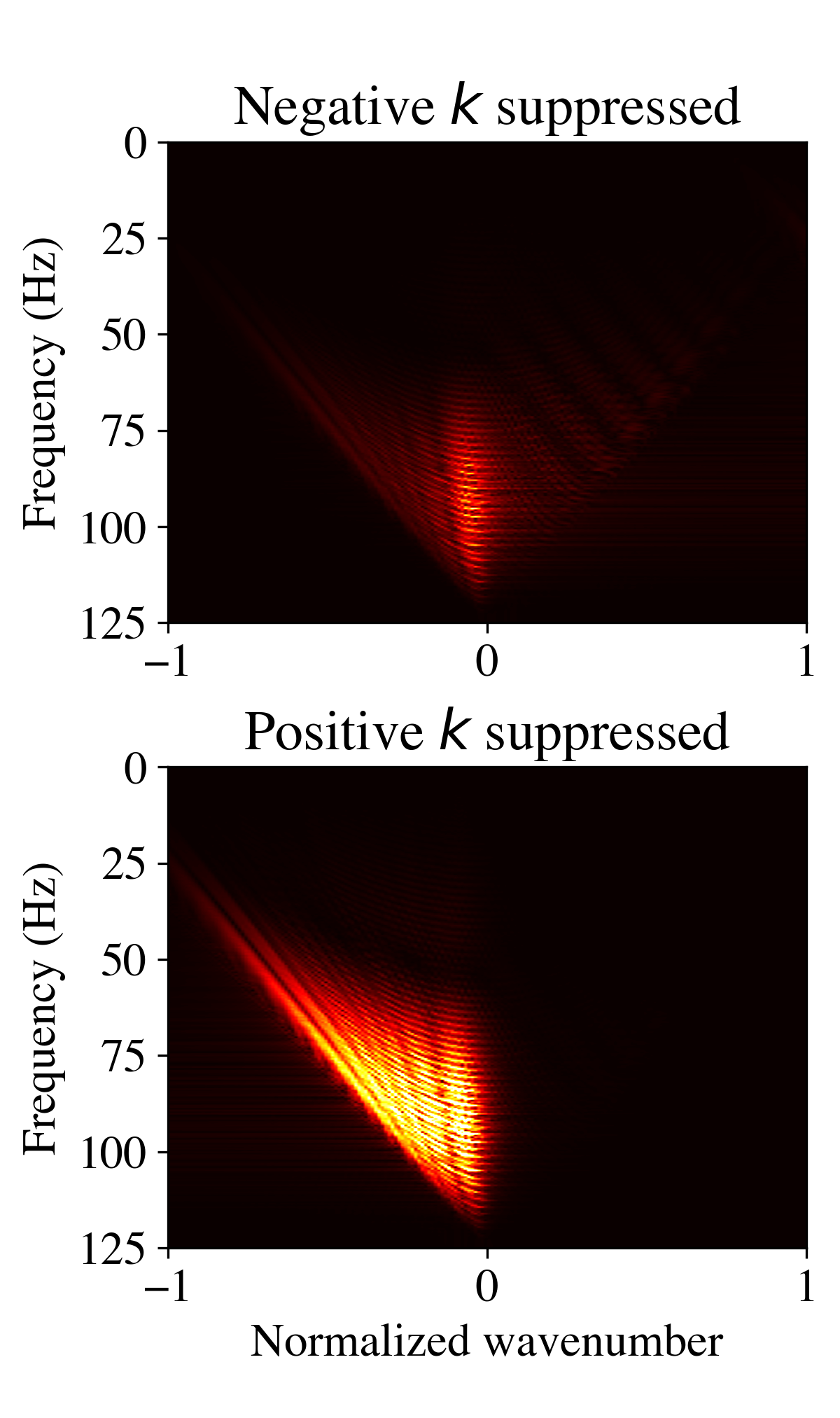} 
	\end{tabular}
	\caption{Examples of using pairs of the proposed transformations on different data and their effect on the data. (a) Synthetic data and two transformed versions using Gamma correction. (b) The effect of Gamma correction on the amplitudes. (c) Real data and two peak-frequency shifted versions. (d) The original and weighted amplitude spectra, alongside the employed weight functions. (e) Synthetic data and two dip-filtered versions. (f) Weighted $f-k$ panels for dip filtering. }
	\label{fig2}
\end{figure}

	\section{Data transformations for ensemble deep learning}
As outlined previously, our network's two branches are fed with differently transformed data. One branch aims to enhance the reconstruction of specific target events, while the other focuses on reconstructing the remaining data. While the second branch may utilize the original data, we employ a transformation to attenuate features amplified in the first branch, enhancing diversity among learners and preventing accuracy loss in other events. We introduce three types of reversible transformations tailored for implementation in the proposed EDM (Figure \ref{fig1}), each designed to target challenging features in the data, including weak events, high frequencies, and crossing dips. These components are known challenges for a basic U-net in self-supervised seismic data reconstruction.

	\subsection{Gamma correction for reconstruction of weak events}
	Gamma correction is an image processing technique used to adjust the contrast of an image by applying a non-linear operation to each pixel. For seismic data ($d$), we use a modified power-law expression:
	
\begin{dmath}
\label{eq5}
	T_\gamma(d)=sign\left(d\right)\left|d\right|^\gamma. 
\end{dmath}

Assuming that $0<|d| \leq 1$,  for $0<\gamma<1$, the transformed data ($T_\gamma({d})$) have a decreased contrast, while for $\gamma>1$, the transformed data have an increased contrast between the strong and weak events (while $0<|T_\gamma({d})|  \leq 1$). A transformation with $0<\gamma<1$ is called \textit{encoding gamma}, and conversely, a transformation with $\gamma>1$ is called \textit{decoding gamma}.

Figure \ref{fig2}a illustrates a synthetic example featuring encoding ($\gamma=0.5$) and decoding ($\gamma=1.25$) transformed versions. Encoding-gamma transformation increases the contribution of weak events in the loss function. Unlike common seismic gain functions, encoding-gamma globally enhances weak events regardless of their location and nearby events. It is also simply reversible using $\frac{1}{\gamma}$ in equation \ref{eq5}, i.e., $T_\gamma^{-1} = T_{\frac{1}{\gamma}}$.

As depicted in Figure \ref{fig2}b, an encoding-gamma transformation extends the range of weaker amplitudes, for example, by mapping ${d} \in (-0.5, 0.5)$ to $(-0.7, 0.7)$. Training a model with this transformed data, the subsequent reverse transformation into the original smaller range effectively mitigates prediction errors associated with weaker events. Conversely, a decoding-gamma transform performs a similar task for higher amplitudes. Therefore, we use both transformations as $T_1$ and $T_2$ (corresponding to $\gamma=0.5$ and $\gamma=1.25$) in our ensemble model so that the two baselines can complete each other for an accurate reconstruction of different ranges of strong and weak events.

\subsection{Peak-frequency shifting for the reconstruction of high-frequencies}

Increasing the relative magnitude of the lower or higher frequencies smooths or sharpens the images, respectively. To shift the peak (dominant) frequency of data samples toward higher or lower frequencies, we multiply their amplitude spectra ($m(f)$) with a Gaussian weight function ($p(f)$). Assuming that $d(t,x)$ is the input data (where $x$ represents the spatial coordinate and $t$ represents time) and $\hat{d}(f,x)$ is its 1D Fourier transform in time, we define:
\begin{dmath}
	\label{eq6}
 T_p(d)=\mathcal{F}^{-1}[\hat{d}(f,x)p(f)], 
\end{dmath}
where $T_p(d)$ is the peak-frequency shifted data, $\mathcal{F}^{-1}$ is the inverse 1D Fourier transform symbol, $f$ is frequency normalized by the Nyquist frequency ($0 \leq f \leq 1$), and
\begin{dmath}
	\label{eq7}
p(f)=\alpha+\frac{1-\alpha^2}{\alpha}e^{-\frac{(f-f_\mu)^2}{2\sigma^2}}, 
\end{dmath}
where $\alpha$ and $f_\mu$ correspond to the relative magnitude and location of the peak, and $\sigma$ determines the width of the Gaussian function. To shift the peak from the current peak frequency $f_{\text{peak}}$ towards $f_\mu$ we define $\alpha=\frac{m(f_\mu)}{m(f_{\text{peak}})}$ and $\sigma=\frac{(f_\mu-f_{\text{peak}})}{4}$. This weight function has one independent parameter $f_\mu$, which is the intended value for the shifted peak in the $f$ vector.

Figure \ref{fig2}c shows a part of real marine data, and its smoothened and sharpened versions obtained by shifting the peak frequency to lower and higher frequencies, which can be used as the two differently transformed data in our EDM. Figure \ref{fig2}d shows the original amplitude spectrum, the shifted versions for enhancing low and high frequencies, and the corresponding weight functions corresponding to equation \ref{eq7}. The proposed weight function in equation \ref{eq7} smoothly changes the amplitudes to move the peak frequency toward a user-defined value of $f_\mu$ while ensuring that the noise at the two ends of the amplitude spectrum is not amplified. The reverse process (to obtain the original from the smoothened or sharpened data) is performed using $\frac{1}{p}$.

\subsection{Dip filtering for the reconstruction of crossing dips}

An $f-k$ transformation is a common seismic data processing technique that decomposes the data into its frequency and wavenumber components. Considering that the wavenumber axis represents the apparent dip of events, an $f-k$ filter can attenuate events based on their dip. We use a smooth weight function ($q$) to suppress the magnitude of events with positive or negative dips in the Fourier domain. Assuming that $d(t,x)$ is the input data and $\hat{d}(f,k)$ is its 2D Fourier transform, we define
\begin{dmath}
\label{eq8}
	T_q(d)=\mathcal{F}_2^{-1}[\hat{d}(f,k)q(f,k)],
\end{dmath}
and
\begin{dmath}
\label{eq9}
	q(f,k)=0.5(1+a+(a-1)\tanh{b(k+c)}),
\end{dmath}
where $T_q(d)$ is the dip-filtered data, $\mathcal{F}_2^{-1}$ represents the inverse 2D Fourier transformation, $k$ is the normalized wavenumber ($-1 \leq k \leq 1$), and $a$ and $b$ determine the minimum weight and the smoothness of weight variation at $c$, respectively. 

Using $c = 0$, the proposed weight function can be used to attenuate positive or negative dips by changing the sign of $b$. Defining $c=\theta\frac{f}{|k|}$, the weight function forms a fan filter used to attenuate events with steeper or gentler dips than $\theta$. Through experiments performed for this study, we selected $a = 0.1$, and $|b| = 8$, and used these values in all the presented results. 

Figure \ref{fig2}e presents a synthetic data sample alongside its versions after transformations for suppression of positive ($b=-8$) and negative ($b=-8$) dips, which can be used as $T_1$ and $T_2$ in our EDM, and Figure \ref{fig2}f shows the corresponding weighted $f-k$ panels. The filtering does not completely remove the opposite dipping events but decreases their magnitudes. As a result, the scarce positive-dipping diffractions become more prominent in the transformed data with suppressed negative dips. These imperfect filters are enough to train each baseline model to reconstruct a different subset of events based on their dips. The filtering is reversible for $a>0$ by using $\frac{1}{q}$.

\section{Training for data reconstruction}
For the reconstruction of missing traces in seismic data through self-supervised learning, we use the target incomplete shot gathers (that have recording gaps) for both training and inference. After selecting the suitable transformations and their hyper-parameters, the training process for each data batch is as follows:

	\begin{enumerate}
		\item{Select a data batch ($d_{gt}$) and compute two transformed representations to obtain $T_1(d_{gt})$ and $T_2(d_{gt})$.}
		\item{Randomly mask a region of consecutive traces within each sample with zeros, designating it as the input $d$ to the network. }
		\item{Train the network to reconstruct the masked traces by minimizing the proposed loss function. The loss is exclusively computed for the masked traces. To ensure the network is not trained to produce zero traces in the output, any masked trace situated over an original gap in the data (where corresponding data are unavailable) is removed from the calculation of loss.}
		\item{Iterate through different selections of masked regions in step 2, enabling the network to learn the continuity of events across all traces.}
	\end{enumerate}
The network, trained to reconstruct randomly selected masked regions from the surrounding data, learns to reproduce the original missing sections of the input data. By random masking-while-training (rather than pre-creating a large number of masked training samples), our method increases the data diversity while maintaining memory efficiency.
	
\begin{figure}[]
	\centering
	\begin{tabular}{@{}p{\linewidth}@{\quad}p{\linewidth}@{}}
		\centering
		\subfigimg[width=.4\linewidth]{(a)}{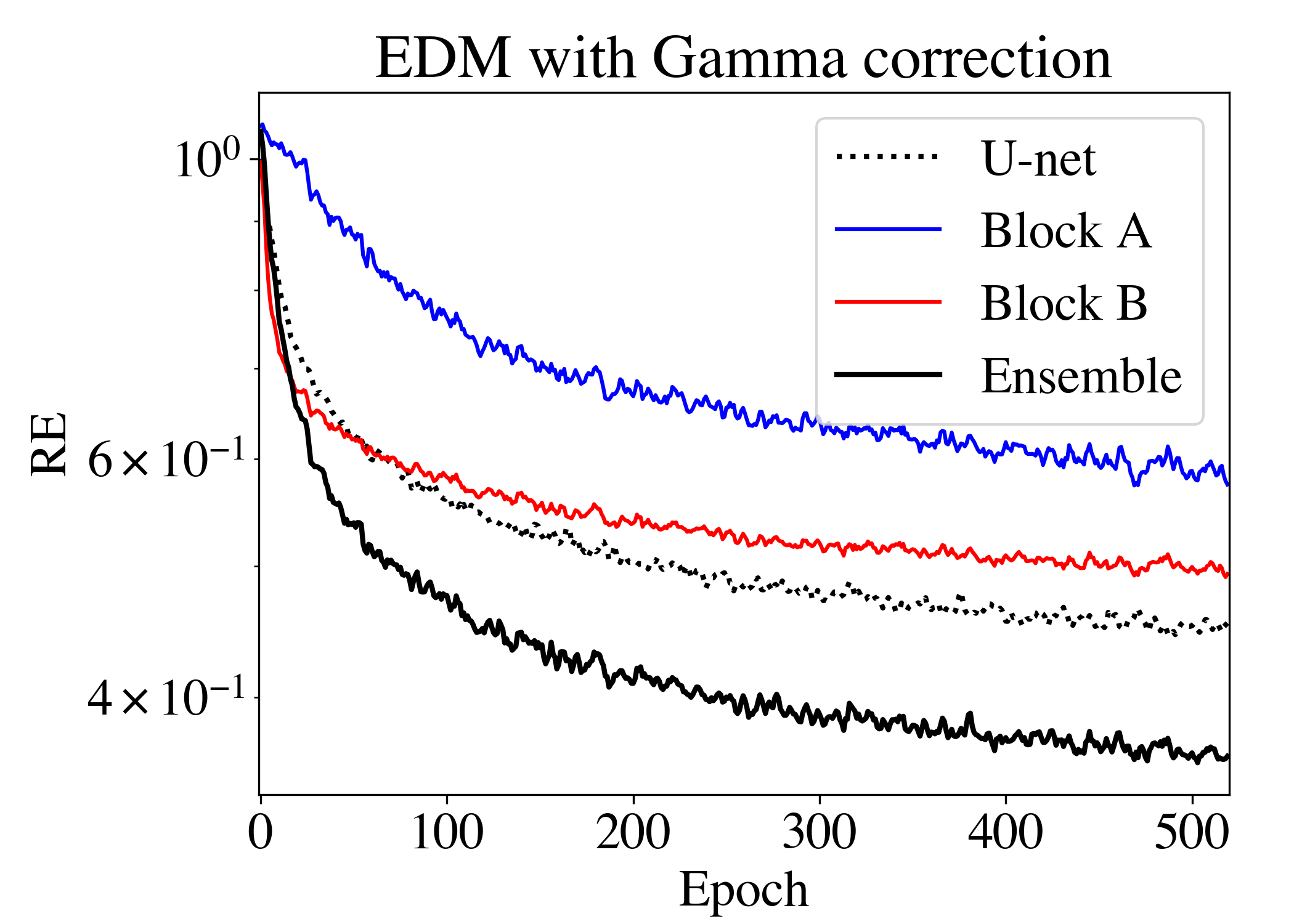} 
		\subfigimg[width=.4\linewidth]{(b)}{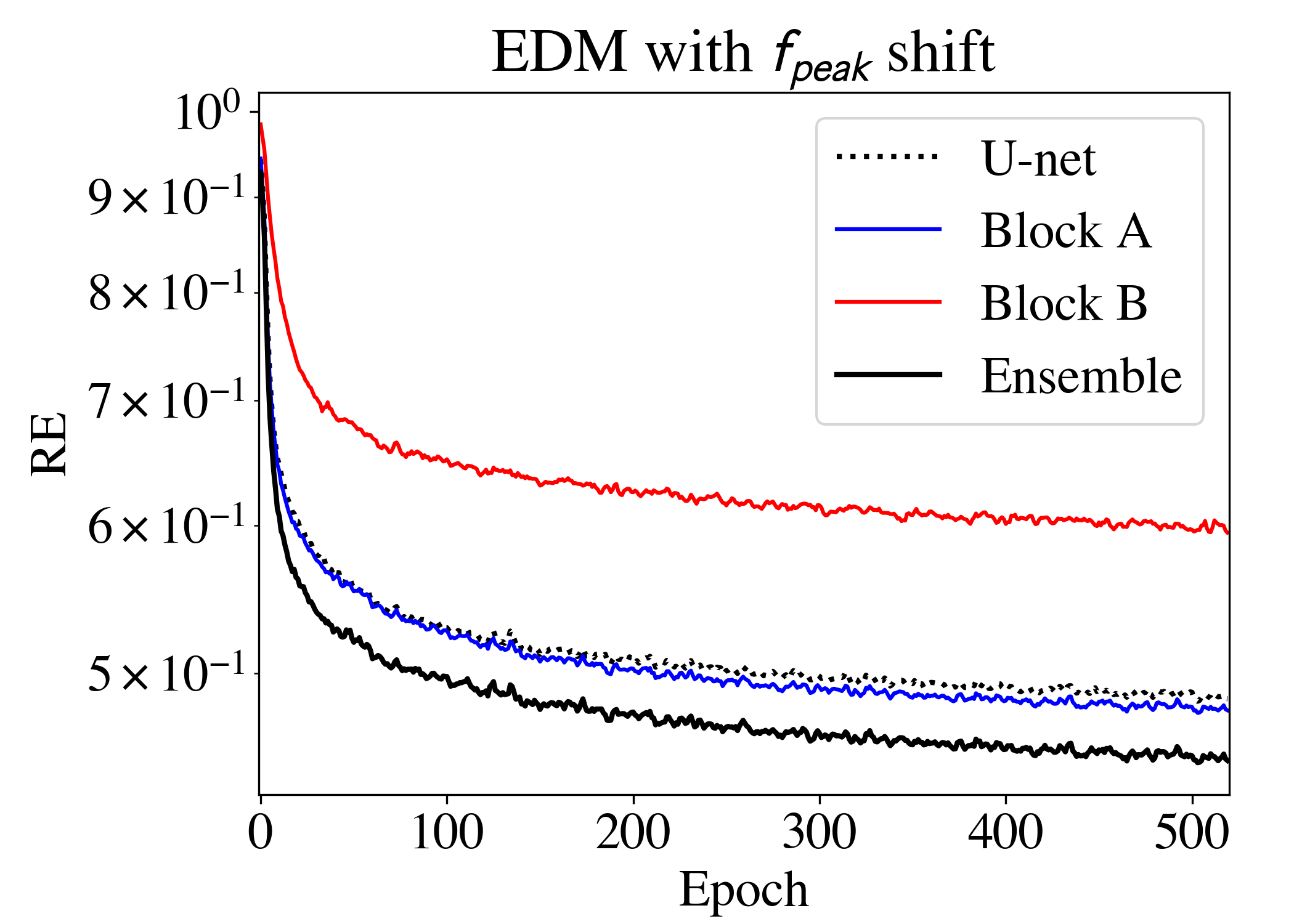} 
		\\
		\subfigimg[width=.4\linewidth]{(c)}{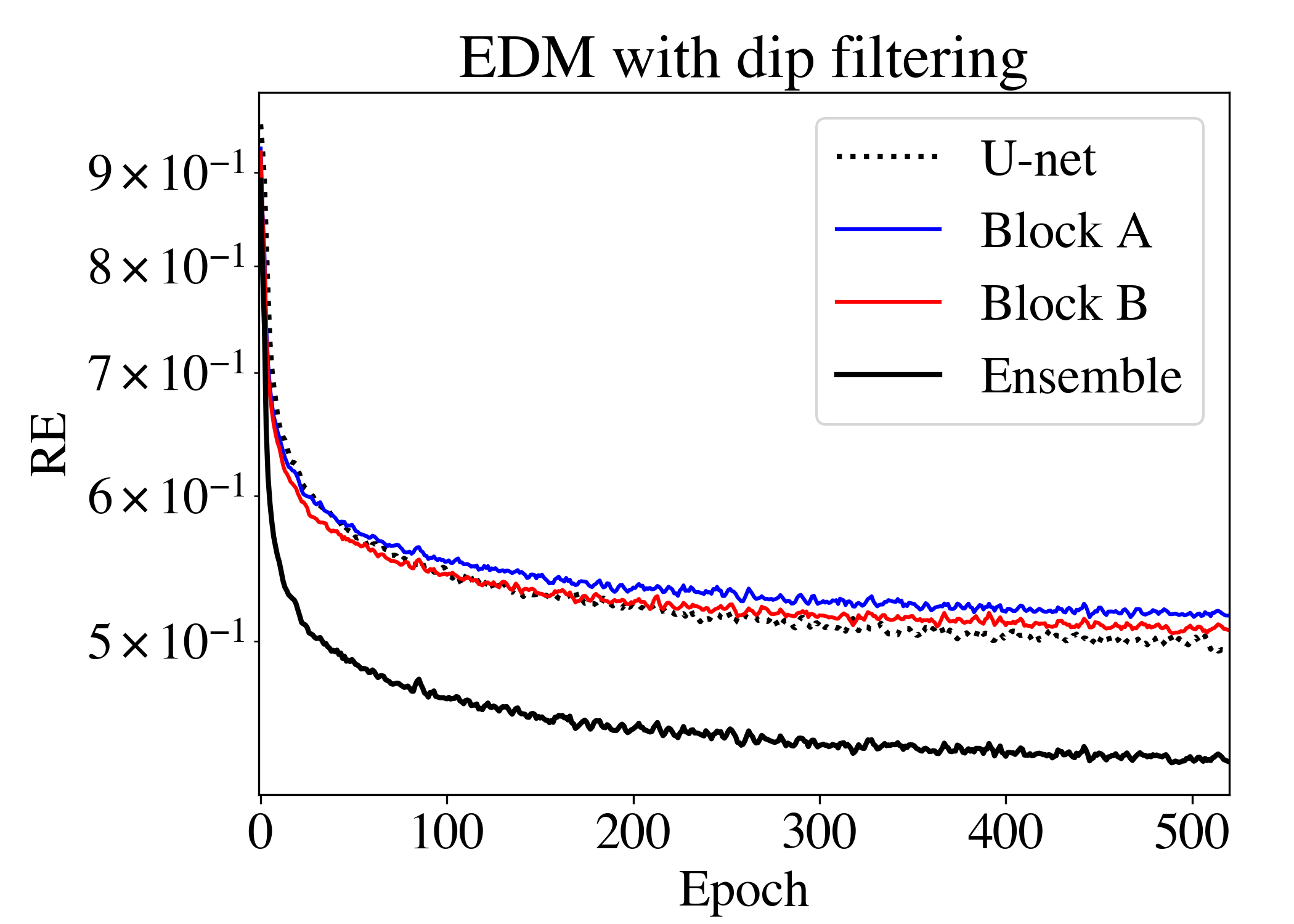} 
		\subfigimg[width=.4\linewidth]{(d)}{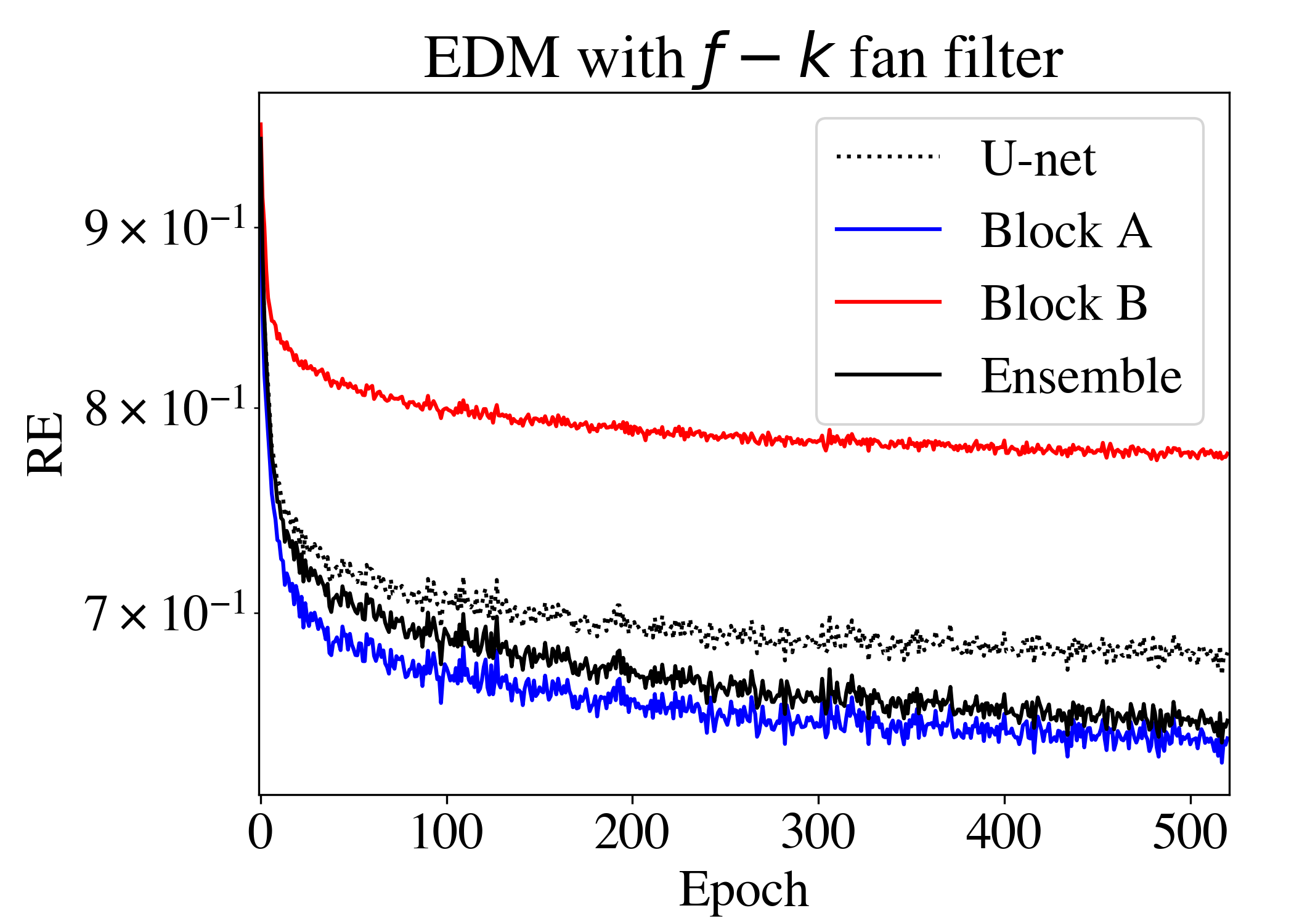} 
	\end{tabular}
	\caption{The evolution of different terms of the loss, the relative mean absolute error (RE) of the reconstructed missing traces in the training data. We conducted four experiments using different data, and different transformations within the ensemble deep model (EDM).}
	\label{fig3}
\end{figure}

\begin{figure}[]
	\centering
	\begin{tabular}{@{}p{\linewidth}@{\quad}p{\linewidth}@{}}
		\centering
		\subfigimg[width=.4\linewidth]{(a)}{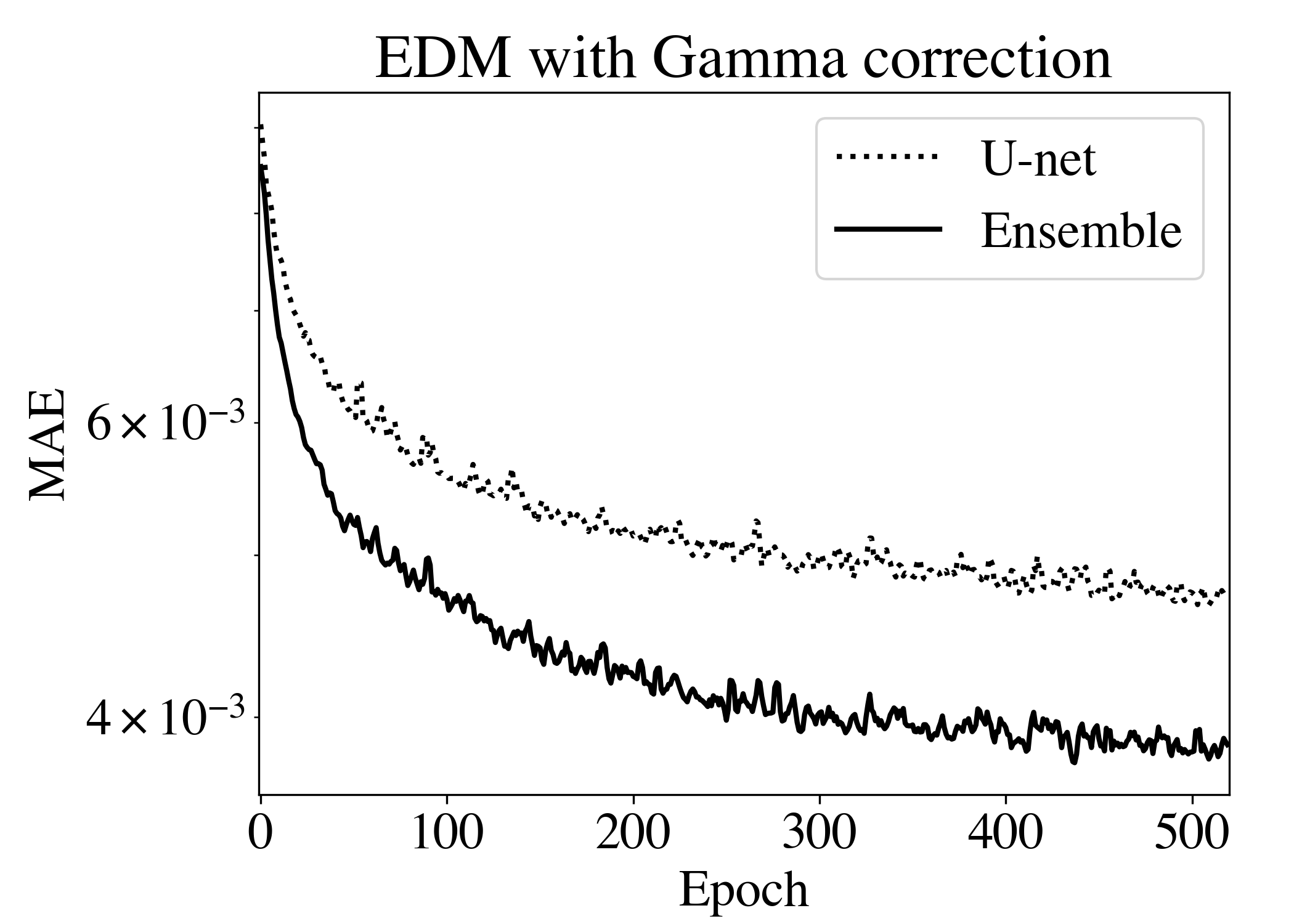} 
		\subfigimg[width=.4\linewidth]{(b)}{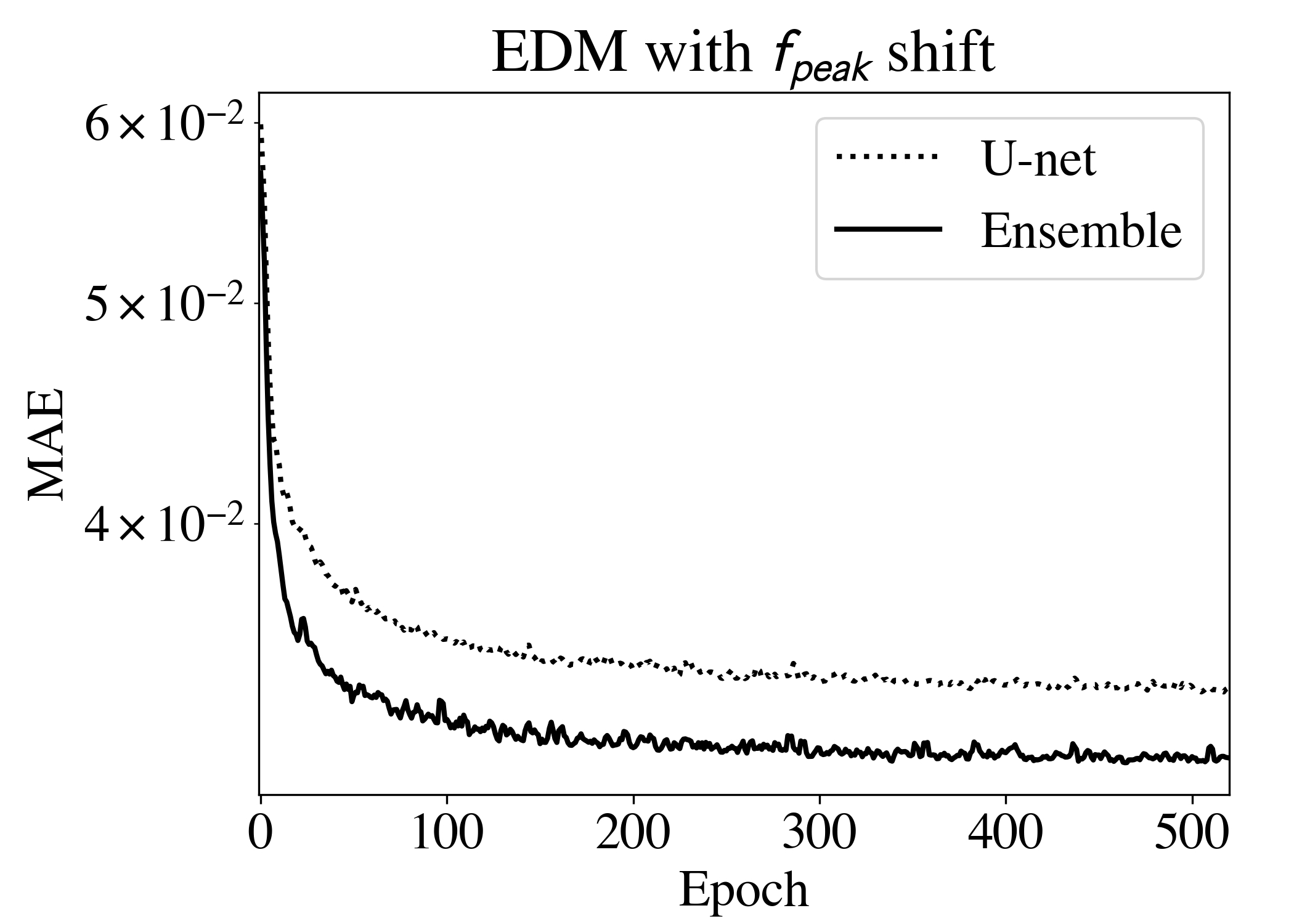} 
		\\
		\subfigimg[width=.4\linewidth]{(c)}{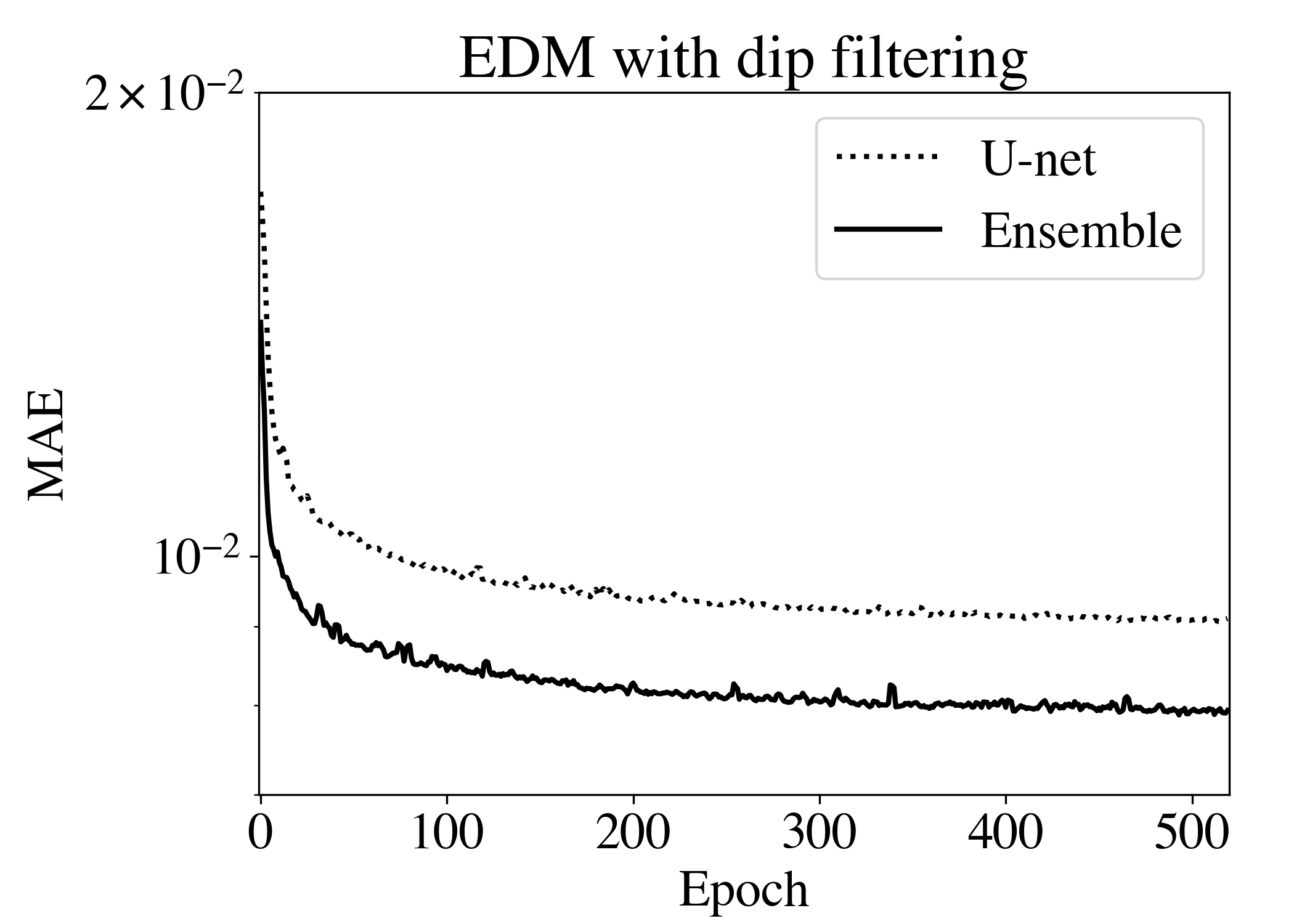} 
		\subfigimg[width=.4\linewidth]{(d)}{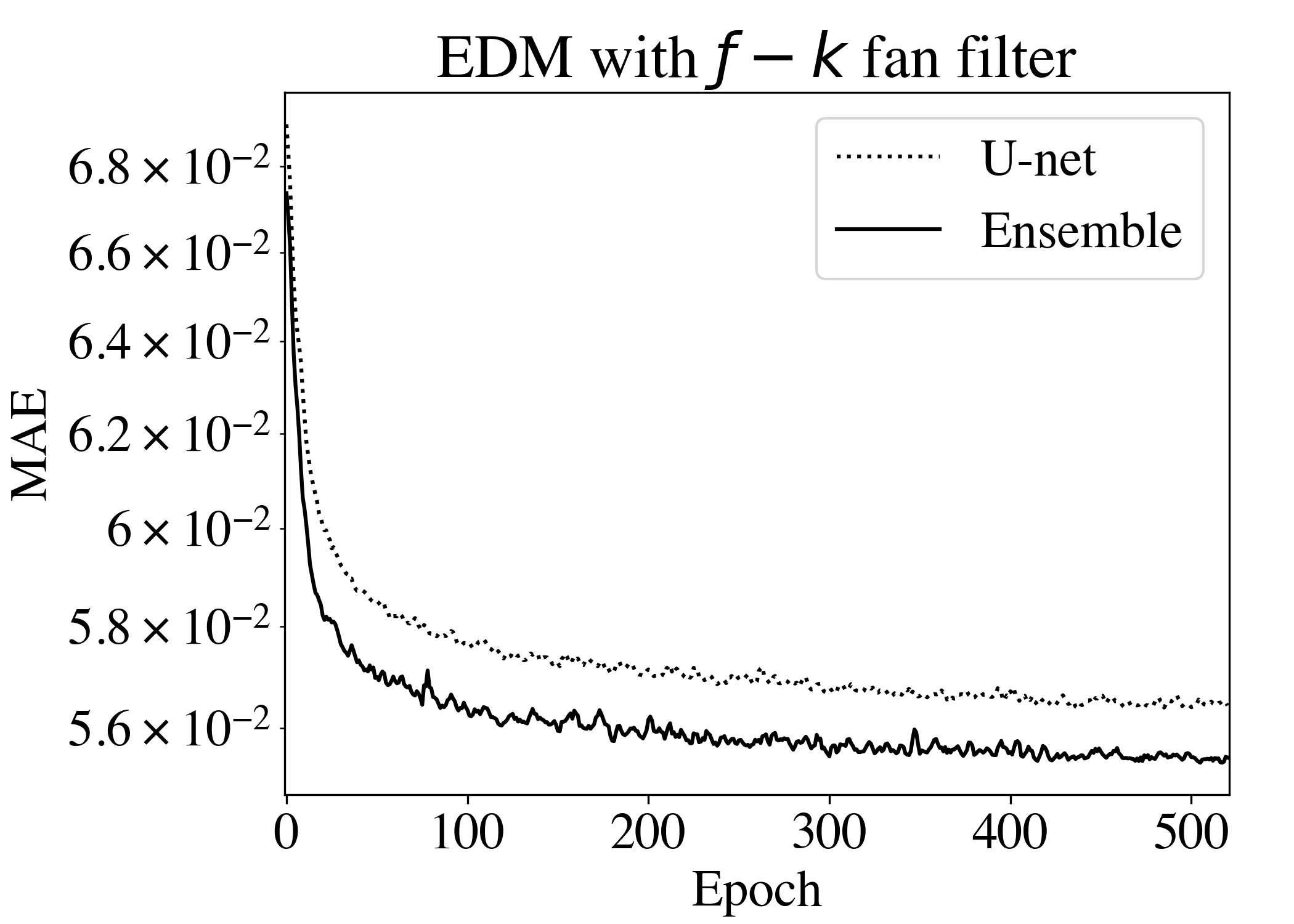} 
	\end{tabular}
	\caption{The evolution of metrics, the mean absolute error (MAE) of the reconstructed missing traces in the test data during training. We conducted four experiments using different data, and different transformations within the ensemble deep model (EDM)}
	\label{fig4}
\end{figure}
	
\section{Applications}
	
In four experiments we apply the proposed method for self-supervised data reconstruction of consecutive missing traces, using two benchmark synthetic and two real datasets. For each experiment, we initially employ a simple U-net and follow the described training procedures. Based on the distinct challenges present in each dataset, we choose one type of the proposed transformations with two sets of hyperparameters for integration into our Ensemble Deep Model (EDM). We then evaluate the accuracy in both training and test data, along with presenting reconstruction examples to compare the performance of our EDM against that of the conventional U-net.

\paragraph{First experiment:} We use a portion of 2D viscoelastic modeled P-wave data from the KFUPM-KAUST Red Sea model (\cite{al2017kfupm}). This dataset comprises 237 shot gathers, each with the dimension of 1152 time samples by 192 traces. A part of a representative shot gather is shown in Figure \ref{fig2}a. We introduce recording gaps of 10 missing traces in all shot gathers and form incomplete data to be used for the proposed training algorithm for data reconstruction. The same number of traces are also used when masking. The network is trained exclusively on incomplete data, and never exposed to complete data during training. We also use random polarity reversal for augmentation. This dataset includes several strong events that deteriorate the reconstruction of crossing weaker events.  

\begin{figure}[]
		\centering
		\includegraphics[width=1\linewidth]{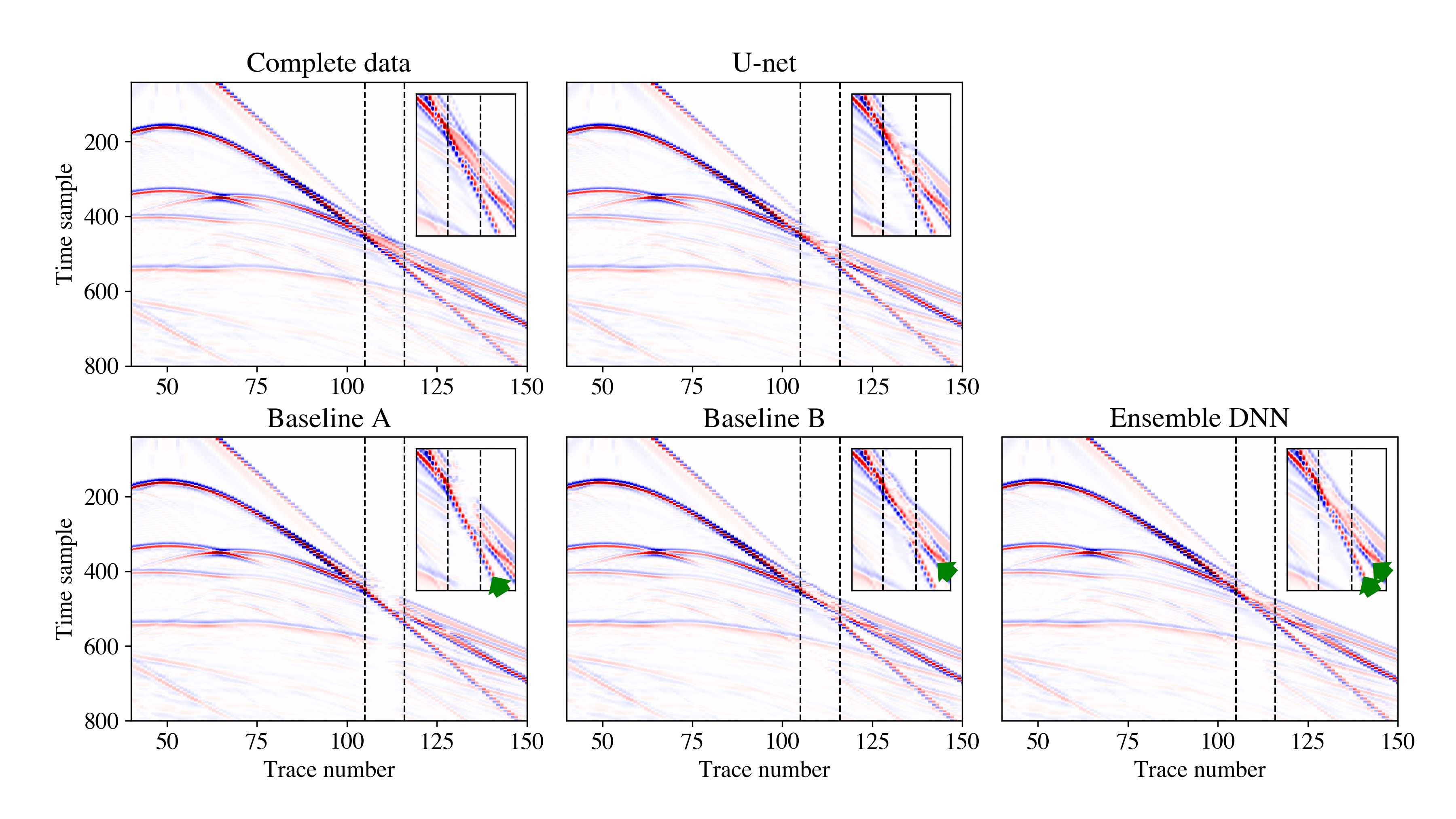}
		\caption{A comparison between the predicted data by our ensemble model incorporating the Gamma correction transformation, and a simple U-net. Two vertical dashed lines mark the reconstructed region. Overlapping windows show an enlarged section. The green arrows show a successful reconstruction of different events in each baseline and their aggregation in the ensemble output. Colormap clipped at $\pm0.5$ for clearer visualization of weaker events.}
		\label{fig5}
	\end{figure}

Initially, we use a conventional U-net (Table 1) and follow the proposed training steps. Figure \ref{fig3}a includes the evolution of the loss, and Figure \ref{fig4}a includes the reconstruction error of originally missing (test) data during training. In Figure \ref{fig5}a part of a shot gather and the reconstructed data with the trained U-net is presented. The network produced high reconstruction errors for weaker events, particularly where they intersect stronger events.

\begin{figure}[]
	\centering
	\includegraphics[width=0.4\linewidth]{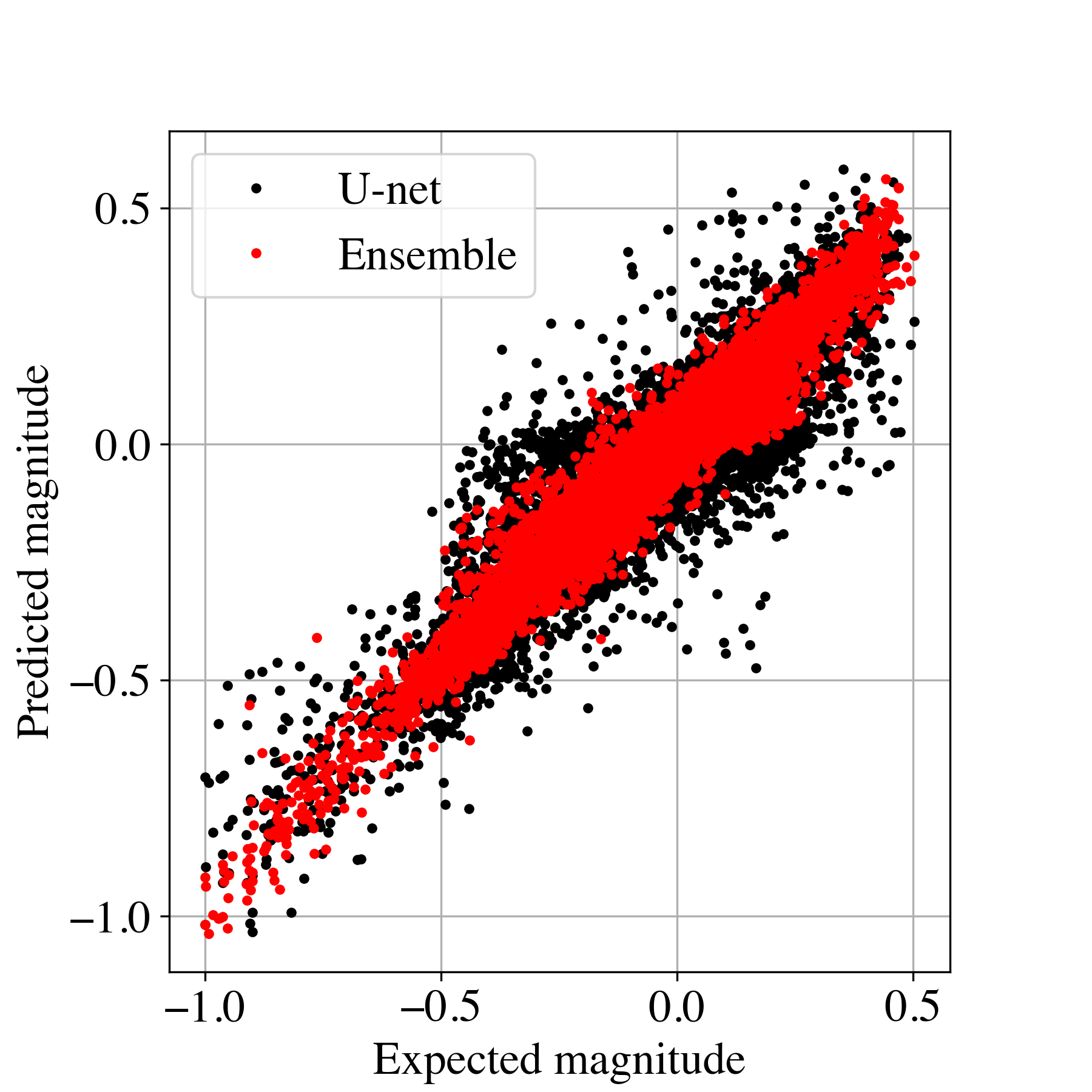}
	\caption{Scattered plot of predicted versus expected amplitudes, related to Figure \ref{fig5}.}
	\label{fig6}
\end{figure}

Subsequently, we construct our EDM using the suggested encoding ($\gamma = 0.5$) and decoding ($\gamma = 1.25$) gamma transforms in two baselines. Examples of the transformed data are shown in Figure \ref{fig2}a. Although the original data contain clearly defined strong events and the problem predominantly impacts weaker events, we opt to use both encoding and decoding transformations. This decision is influenced by our observation of better performance compared to using the original data ($\gamma = 1$) in one baseline. This phenomenon can be attributed to the increased diversity of models and the prevention of accuracy degradation in strong events, as described in the Gamma correction subsection. Figure \ref{fig3}a depicts the successful minimization of different terms of the proposed loss function during training.  As Figures \ref{fig3}a and \ref{fig4}a show that the overall reconstruction error of our EDM remains lower than the U-net for the training and the test missing data, respectively. The example in Figure \ref{fig5} shows that the trained EDM separately reconstructed strong and weak events in two baselines, resulting in an accurate reconstruction of both events. Figure \ref{fig6} shows a cross-plot of the expected versus the predicted magnitudes from the trained U-net and EDM. Our EDM improved the accuracy, especially for weaker events. 

\paragraph{Second experiment:} We use a vintage real dataset recorded in the Gulf of Mexico. This dataset consists of 300 shot gathers, each with dimensions of 704 time samples by 176 traces. A portion of a representative shot gather from this dataset is depicted in Figure \ref{fig2}c. We introduce recording gaps of 10 missing traces in all shot gathers and use the incomplete data to train different networks. We also use random polarity reversal for data augmentation. Being real data, the dataset presents horizontal inconsistencies that pose challenges for reconstruction, particularly over large gaps and higher frequencies.

\begin{figure*}[]
		\centering
		\includegraphics[width=1\linewidth]{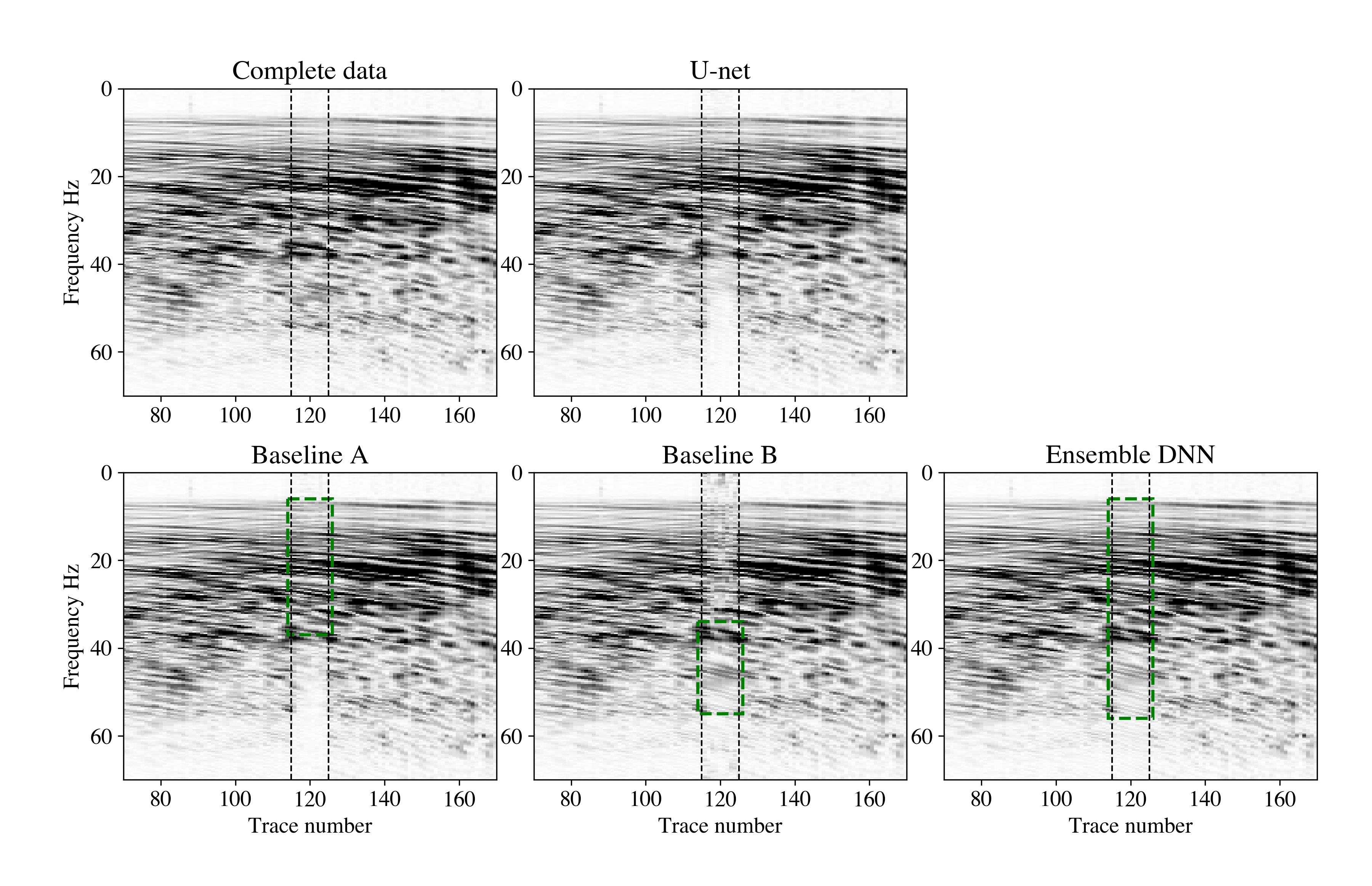}
		\caption{A comparison between our ensemble model with peak-frequency shifting and a simple U-net. Dashed lines mark the reconstructed region. Green boxes indicate the successful reconstruction of low and high frequencies in each baseline and their combination in the ensemble output.}
		\label{fig7}
\end{figure*}

Initially, we train the conventional U-net (Table 1) following the proposed training steps. Figure \ref{fig3}b includes the evolution of the loss, and Figure \ref{fig4}b includes the evolution of the reconstruction error of the test missing data during training. Training continues until no meaningful improvements are observed. Figure \ref{fig7} presents a part of the shot gather and the reconstructed data in frequency-offset space. The U-net is unable accurately reconstruct the higher frequencies in the missing data. 

Subsequently, we build our EDM using the proposed peak-frequency shift toward higher ($f_\mu=0.4$) and lower frequencies ($f_\mu=0.15$)  in two baselines. These values are selected considering $f_{\text{peak}}=0.19$. Examples of the transformed data are shown in Figure \ref{fig2}c. As Figure \ref{fig3}b shows for the training, as well as Figure \ref{fig4}b shows for the test missing data, the overall reconstruction error of our EDM remains consistently lower than the U-net. Moreover, as depicted in the example presented in Figure \ref{fig7}, the trained EDM effectively combines the most accurate reconstruction from each baseline, highlighting its enhanced performance.

\paragraph{Third experiment:} We utilize a portion of the 2004 BP benchmark synthetic data (\cite{billette2005}). This dataset is modeled based on the geologic features and challenges in different sites including the Gulf of Mexico, the Caspian Sea, and the North Sea, offering a comprehensive representation of real-world models. The dataset comprises 948 shot gathers, each with 576 time samples and 192 traces. A part of a representative shot gather is illustrated in Figure \ref{fig2}e. We introduce recording gaps of 20 missing traces across various shot gathers to form the incomplete data subject to data reconstruction. During training, the masking is also performed with 20 subsequent traces. To augment the dataset, we incorporated random polarity reversal and random horizontal flipping. Notably, this dataset encompasses diffractions with distinct shapes and dipping compared to those of the dominant reflections.

\begin{figure*}[]
		\centering
		\includegraphics[width=1\linewidth]{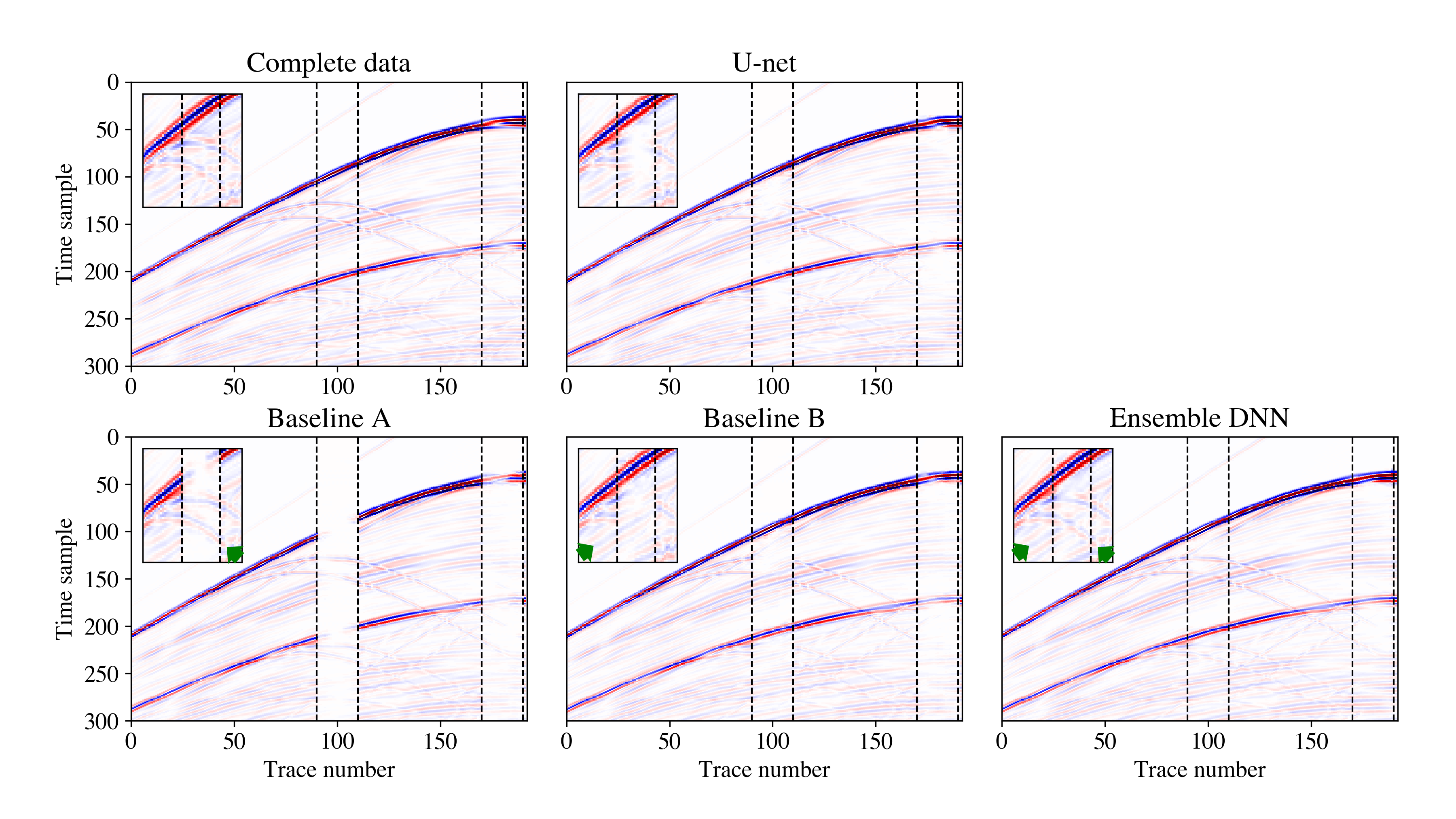}
		\caption{A comparison between our ensemble model incorporating dip-filtering for positive and negative dips and a simple U-net. Dashed lines mark the reconstructed region. Green arrows indicate the successful reconstruction of positive and negative dipping events in each baseline and their combination in the ensemble output. Colormap is clipped at $\pm0.5$ for clearer visualization of weaker events.}
		\label{fig8}
\end{figure*}

Initially, we trained the conventional U-net (refer to Table 1) following the prescribed training steps. Figure \ref{fig4}b illustrates the evolution of the reconstruction error for the originally missing (test) data during training. In Figure \ref{fig8} a part of a shot gather and the reconstructed data of two adjacent gaps is presented. Despite employing the augmented data, the U-net struggles to accurately reconstruct refractions, particularly when their direction contradicts the dominant reflection events.

Subsequently, we build our EDM using the proposed dip filtering for positive ($a = 0.1$, $b = 8$, $c = 0$) and negative ($a = 0.1$, $b = -8$, $c = 0$) dips in two baselines. Examples of the transformed data are shown in Figure \ref{fig2}e. As Figure \ref{fig4}c shows, the overall reconstruction error of our EDM remains lower than the U-net during training. While the intended enhancement primarily targets scarce diffractions, the reduction in average error may be insignificant. However, as the example in Figure \ref{fig8} shows the trained EDM separately reconstructed the positive and negative dipping events in two baselines, leading to an accurate reconstruction of both events. 

\paragraph{Fourth experiment:} In the final test, we use real land data recorded in a mountainous area that contains strong groundroll and other types of noise. The dataset comprises 300 shot gathers, each with the dimension of 620 time samples by 352 traces. We create gaps in the data by muting the recordings of up to six consecutive receivers. We repeat the aforementioned process of initially training a U-net, and then building and training our ensemble model with the suggested dip filtering. We define the parameters of the suggested dip filtering for a smooth $f-k$ fan filtering ($a = 0.1$, $b = \pm8$, $c = 1\frac{f}{k}$). Using two signs of $b$ in two baselines divide the reconstruction tasks of the steeper events (e.g., groundroll) and flatter events (e.g., reflections) between baselines.

\begin{figure*}[]
	\centering
	\includegraphics[width=1\linewidth]{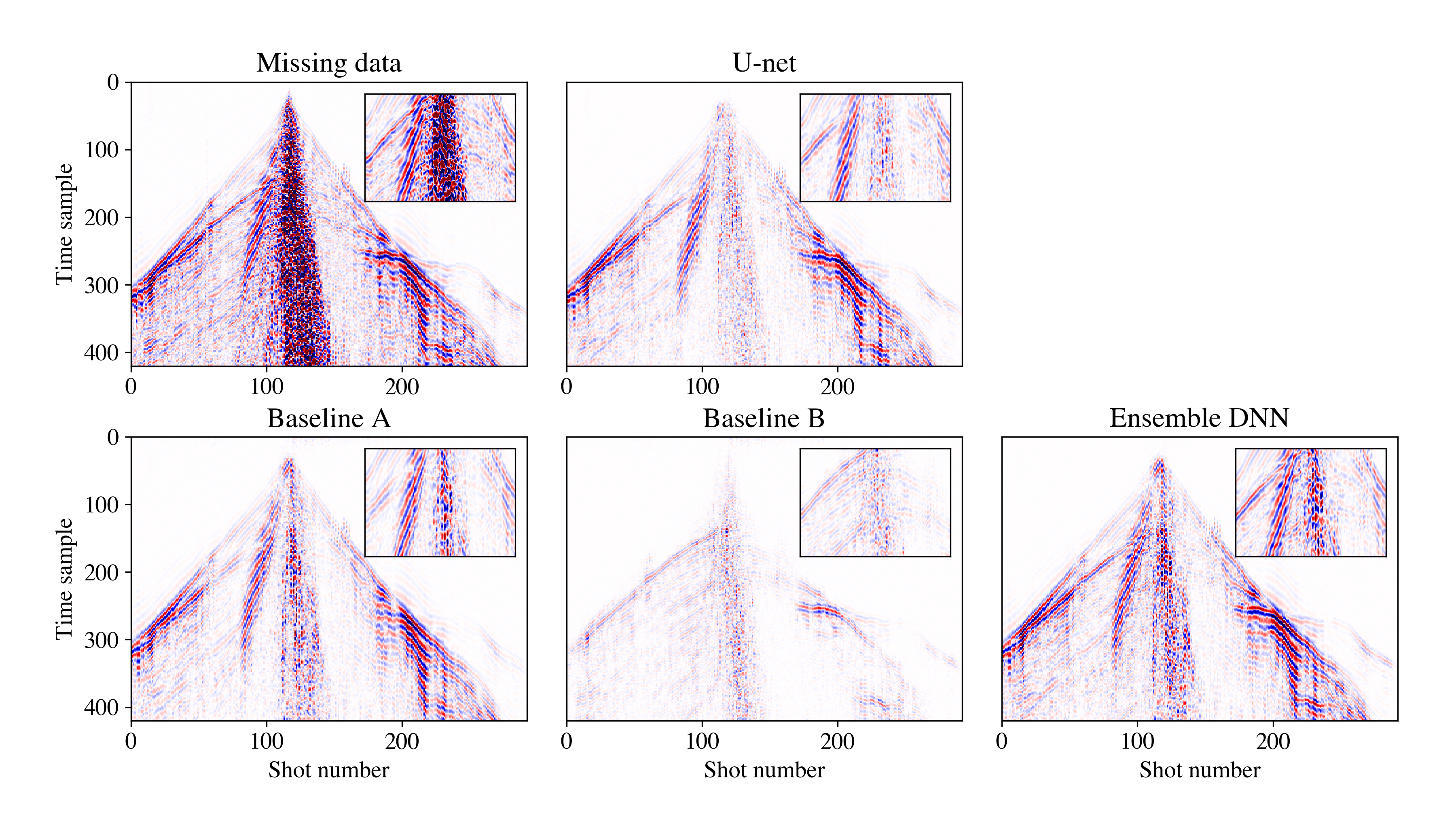}
	\caption{A comparison between our ensemble model incorporating $f-k$ fan filtering and a simple U-net. The entire data are reconstructed CRGs from 300 shot gathers. Colormap is clipped at $\pm0.5$ for clearer visualization of weaker events.}
	\label{fig9}
\end{figure*}

\begin{figure*}[]
	\centering
	\includegraphics[width=0.9\linewidth]{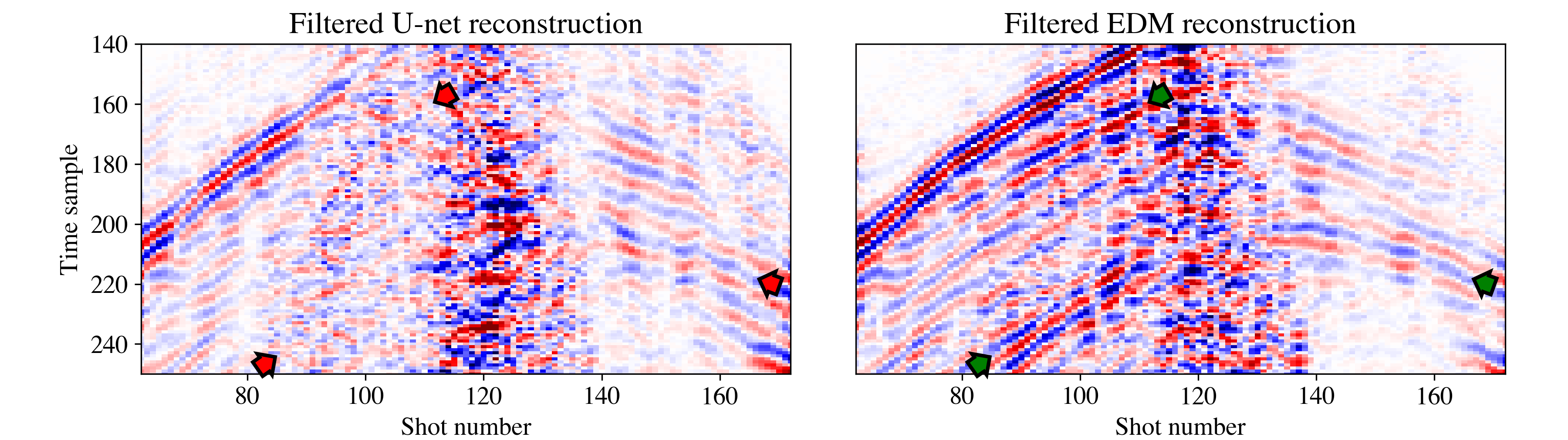}
	\caption{Filtered reconstruction results of two networks show the signal beneath the groundroll. Green arrows indicate the successful reconstruction of reflections in the ensemble output and the red arrows mark the same locations in the U-net output.}
	\label{fig10}
\end{figure*}

Figure \ref{fig3}d shows the evolution of different terms of the loss, and Figure \ref{fig4}d shows the evolution of the reconstruction error of the test missing data during training. Due to the presence of irregularities and incoherent noise, the reconstruction errors are higher than the previous tests, but the proposed ensemble model outperforms the simple U-net in terms of accuracy. Figure \ref{fig9} shows the reconstruction of a common receiver gather (CRG) in the middle of a gap of six traces (The entire CRG is reconstructed from the corresponding shot gathers). Each baseline has successfully reconstructed different events, resulting in the reconstruction of overlapping events by the ensemble model. To highlight this feature, we filter the groundroll from the reconstruction result of the U-net and ensemble models and show a zoomed section in Figure \ref{fig10}. Unlike the simple U-net, the ensemble model has successfully reconstructed the reflection events under the groundroll.

	\section{Discussion}
The proposed ensemble deep learning approach offers distinct training for each of the baselines, using a pair of forward and reverse transformations with distinct parameters. Although, in general, the complicated forward and reverse transformations can also be learned by the neural netwroks, incorporating the known transformations as explicit modules inside the architecture allows us to incorporate our prior knowledge of the data into the network, offering better performance on limited datasets.  In addition, the proposed method also offers a unique advantage in terms of explainability. By design, our EDM incorporates distinct baselines tailored to specific tasks, providing a form of a priori explainability. This characteristic allows for a deeper understanding of the model's inner workings, enabling an interpretation of the contributions of individual branches. This inherent explainability can enhance our trust in the model's predictions, and reduce concerns related to the general black-box nature of neural networks.

We suggested three types of transformations, each to address the observed limitations of conventional networks in handling data with distinct features. However, the approach can be used with different types of transformations and can be extended to more than two baselines to handle more complex data. Exploring alternative methods like the Radon and wavelet transforms and inverse data space manipulation, which can enhance or suppress specific data features, fits the presented framework but requires further investigation. A requirement is for the transformation to have an inverse i.e., it should be bijective. For example, the presented Gamma correction in equation \ref{eq5} is reversible for $\gamma>0$.

The weights in the loss function balance the contributions of the baselines and the ensemble outputs, ensuring that the ensemble effectively uses the outputs of diverse branches while maintaining overall coherence in the reconstruction. Assigning a higher weight to the ensemble output compared to the outputs of each baseline aligns with the main objective of the ensemble model, which aims to enhance overall reconstruction performance by leveraging the complementary strengths of multiple models. However, through several experiments, we observed limited sensitivity to the value of weights and decided on a fixed value of $\lambda_j=0.5$ in the presented examples. This may be attributed to the collaborative nature of the baselines and the ensemble, which collectively work towards the common objective of data reconstruction.

The suggested data transformations offer adaptability to different datasets and applications using the included hyperparameters. This hyperparameter adjustment represents an additional cost compared to simpler models. To avoid a higher cost, we introduce weight functions within the transformations using minimal adjustable parameters. Another consideration is the increased training time associated with our method. Across all applications presented, our approach required roughly three times the training time compared to using a basic U-net. In many applications, the improvement in accuracy may justify the additional cost, but there may be room to improve the efficiency of the approach, which will be the subject of future study.

\section{Conclusion}
In scenarios where conventional models struggle to capture specific data details, our proposed Ensemble Deep Model (EDM) offers a promising solution by allocating dedicated components of the model to learn these features. By integrating a pair of data transformations within the loss and architecture, our method enhances learning by amplifying specific features in one baseline while suppressing them in the other so that they complement each other's tasks effectively. The explicit inclusion of data transformation modules, as opposed to relying solely on the network to learn them, makes our approach more suitable for applications with small datasets typically encountered in self-supervised methods. Based on comparative analyses against a simple U-net using two synthetic and two real datasets, we demonstrated the superior accuracy of our EDM in reconstructing events with varying magnitudes, frequencies, and complexities. This includes scenarios containing weak and strong events, high and low frequencies, and overlapping events such as diffractions and reflections covered by groundroll. Despite the superior performance, our method requires new hyper-parameter adjustment and about three times the training cost of a simple U-net.
	
	\section*{Acknowledgments}
	This work is supported by: the Spanish Ministry of Science and Innovation projects with references TED2021-132783B-I00 funded by MCIN/ AEI /10.13039/501100011033 and the European Union NextGenerationEU/ PRTR, PID2019-108111RB-I00 funded by MCIN/ AEI /10.13039/501100011033, the “BCAM Severo Ochoa” accreditation of excellence CEX2021-001142-S / MICIN / AEI / 10.13039/501100011033; the Spanish Ministry of Economic and Digital Transformation with Misiones Project IA4TES (MIA.2021.M04.008 / NextGenerationEU PRTR); and the Basque Government through the BERC 2022-2025 program, the Elkartek project BEREZ-IA (KK-2023/00012), and the Consolidated Research Group MATHMODE (IT1456-22) given by the Department of Education. Tariq Alkhalifah thanks KAUST for its support. We acknowledge BP,  KFUPM, and KAUST for making the synthetic data available.
	
	\section*{Data Availability}
	A Python implementation for the proposed ensemble deep learning for data reconstruction can be accessed at: https://github.com/mahdiabedi. 
	Real data in this study are confidential and cannot be shared.
	The authors report no conflict of interest.
	
	\bibliography{references_Ensemble_Deeplearning}

\end{document}